\documentclass{article}%
\usepackage{graphicx}
\usepackage{amsmath}
\usepackage{amsfonts}
\usepackage{upgreek}
\usepackage{amssymb}%
\providecommand{\U}[1]{\protect\rule{.1in}{.1in}}

\begin{document}

\author{Antony Valentini\\Augustus College}

\begin{center}
{\LARGE De Broglie-Bohm Quantum Mechanics}

\bigskip

Antony Valentini\footnote{email: a.valentini@imperial.ac.uk}

\textit{Theoretical Physics, Blackett Laboratory, Imperial College London,}

\textit{Prince Consort Road, London SW7 2AZ, United Kingdom.}

\bigskip
\end{center}

We provide an overview of the de Broglie-Bohm pilot-wave formulation of
quantum mechanics, emphasising its applications to field theory, high-energy
physics, gravitation, and cosmology.

\bigskip

1 Introduction

2 Low-energy many-body systems

3 Quantum `measurements'

4 Quantum relaxation

5 Beyond quantum mechanics

\qquad{\small 5.1 Nonlocality and nonequilibrium signalling}

\qquad{\small 5.2 Breaking uncertainty}

\qquad{\small 5.3 Subquantum measurement}

\qquad{\small 5.4 Subquantum information and subquantum computation}

6 Quantum field theory and high-energy physics

\qquad{\small 6.1 Scalar field}

\qquad\qquad{\small 6.1.1 Emergent quantum theory and Lorentz invariance}

\qquad\qquad{\small 6.1.2 Remarks on Lorentz and Galilean invariance}

\qquad\qquad{\small 6.1.3 Quantum field relaxation}

\qquad{\small 6.2 Fermions}

\qquad\qquad{\small 6.2.1 Grassmann field theory}

\qquad\qquad{\small 6.2.2 Dirac sea theory}

\qquad{\small 6.3 Electromagnetic field. Scalar QED}

\qquad{\small 6.4 Non-Abelian gauge theories}

\qquad\qquad{\small 6.4.1 QCD}

\qquad\qquad{\small 6.4.2 Electroweak Theory. Spontaneous Symmetry Breaking}

7 The early universe

\qquad{\small 7.1 Quantum fields in curved spacetime}

\qquad{\small 7.2 Suppression of quantum relaxation on expanding space}

8 Quantum gravity and quantum cosmology

\qquad{\small 8.1 Pilot-wave theory and the Wheeler-DeWitt equation}

\qquad{\small 8.2 Dynamical consistency. Preferred foliation}

\qquad{\small 8.3 Pilot-wave quantum cosmology}

\qquad{\small 8.4 Problem of probability. Beyond the Born rule}

9 Conclusion

Bibliography

\bigskip

To appear in: \textit{Encyclopedia of Mathematical Physics, 2nd edition},
Amsterdam: Elsevier.

\section{Introduction}

The de Broglie-Bohm pilot-wave formulation of quantum mechanics was first
proposed by de Broglie in 1927, then revived and extended by Bohm in 1952 (de
Broglie 1928; Bohm 1952a,b; Holland 1993; Bacciagaluppi and Valentini 2009;
Valentini 2025). It is a deterministic dynamics of trajectories for individual
quantum systems, where each trajectory is guided by a physical `pilot wave' in
configuration space. The theory is nonlocal, as required by Bell's theorem,
and it resolves the quantum measurement problem (Bell 1987). For ensembles of
systems, it reproduces the statistical predictions of quantum mechanics if the
initial conditions obey the Born rule (Bohm 1952a,b). For more general initial
conditions, the Born rule is violated and we obtain new physics outside the
domain of quantum mechanics (Valentini 1991a,b, 1992). The validity of the
Born rule today can be explained by a past process of dynamical relaxation to
`quantum equilibrium'. A more general nonequilibrium physics may, however,
have existed in the early universe, and could exist today in some exotic
circumstances. According to pilot-wave theory, quantum physics is only a
special case of a wider physics which should be accessible at least in principle.

Employing units $\hbar=1$, consider a general system with configuration-space
wave function $\psi(q,t)$. The Schr\"{o}dinger equation%
\begin{equation}
i\frac{\partial\psi}{\partial t}=\hat{H}\psi\label{Sch_gen}%
\end{equation}
implies the continuity equation%
\begin{equation}
\frac{\partial\left\vert \psi\right\vert ^{2}}{\partial t}+\partial_{q}\cdot
j=0\ , \label{Cont_psi2_gen}%
\end{equation}
where $\partial_{q}\cdot j=2\operatorname{Re}\left(  i\psi^{\ast}\hat{H}%
\psi\right)  $ and $\partial_{q}$ is a gradient in configuration space. For a
given Hamiltonian $\hat{H}$, we can write $j=j\left[  \psi\right]  =j(q,t)$ in
terms of $\psi$ (where $j$ is determined up to a divergence-free term)
(Struyve and Valentini 2009). We can then postulate the de Broglie equation of
motion%
\begin{equation}
\frac{dq}{dt}=v(q,t)=\frac{j(q,t)}{|\psi(q,t)|^{2}} \label{deB_gen}%
\end{equation}
for individual trajectories $q(t)$. In this theory, $\psi$ is a physical field
on configuration space guiding the motion of an individual system. For a given
wave function $\psi(q,t)$, the actual trajectory $q(t)$ is determined by the
initial condition $q(0)$ via the first-order equation of motion (\ref{deB_gen}).

Now consider an ensemble of systems with the same wave function $\psi$. The
systems follow the velocity field $v$, hence a distribution $\rho(q,t)$ of
configurations satisfies the continuity equation%
\begin{equation}
\frac{\partial\rho}{\partial t}+\partial_{q}\cdot\left(  \rho v\right)  =0\ .
\label{Cont_rho_gen}%
\end{equation}
This takes the same form as (\ref{Cont_psi2_gen}) or%
\begin{equation}
\frac{\partial\left\vert \psi\right\vert ^{2}}{\partial t}+\partial_{q}%
\cdot\left(  |\psi|^{2}v\right)  =0 \label{Cont_psi2_gen2}%
\end{equation}
(with $j=|\psi|^{2}v$). It follows that if $\rho=\left\vert \psi\right\vert
^{2}$ initially then $\rho=\left\vert \psi\right\vert ^{2}$ at all times. This
is the state of `quantum equilibrium', with probability density given by the
Born rule. In this state, the theory yields the same statistical predictions
as quantum mechanics (Bohm 1952a,b).

In principle, however, there can exist initial `quantum nonequilibrium'
ensembles with $\rho\neq\left\vert \psi\right\vert ^{2}$ (Valentini 1991a,b,
1992). Such ensembles violate the statistical predictions of quantum
mechanics. However, in appropriate conditions, such ensembles relax to the
equilibrium state $\rho=\left\vert \psi\right\vert ^{2}$, a process that may
have taken place in the early universe, explaining why we observe the Born
rule today (Valentini and Westman 2005). Even so, there could exist residual
nonequilibrium effects, for example in relic particles from the early universe
(Valentini 2007; Underwood and Valentini 2015).

The Schr\"{o}dinger equation (\ref{Sch_gen}) is a field equation on
configuration space. It can be derived from a Lagrangian $L=\int
\mathcal{L}\ dq$ where%
\begin{equation}
\mathcal{L}=\frac{i}{2}\left(  \psi^{\ast}\dot{\psi}-\psi\dot{\psi}^{\ast
}\right)  -\frac{1}{2}\psi^{\ast}\hat{H}\psi-\frac{1}{2}\psi\hat{H}\psi^{\ast}%
\end{equation}
is invariant under global phase transformations%
\begin{equation}
\psi\rightarrow\psi e^{i\theta} \label{global}%
\end{equation}
(where $\theta$ is an arbitrary constant). Noether's theorem then implies a
local conservation law, which coincides with (\ref{Cont_psi2_gen}) (Struyve
and Valentini 2009). Thus the de Broglie velocity (\ref{deB_gen}) originates
from the global symmetry (\ref{global}) of $\mathcal{L}$.

We have assumed for simplicity that $\psi$ has only one component. For a
system with spin, $\psi$ can have more than one component. The theory is
easily generalised to such cases.

\section{Low-energy many-body systems}

We now consider a system of low-energy spinless particles. The wave function
$\psi=\psi(\mathbf{x}_{1},\mathbf{x}_{2},...,\mathbf{x}_{N},t)$ obeys the
Schr\"{o}dinger equation%
\begin{equation}
i\frac{\partial\psi}{\partial t}=-\sum_{n=1}^{N}\frac{1}{2m_{n}}\nabla_{n}%
^{2}\psi+V\psi\ , \label{Sch_Npart}%
\end{equation}
which implies a continuity equation%
\begin{equation}
\frac{\partial\left\vert \psi\right\vert ^{2}}{\partial t}+\sum_{n=1}%
^{N}\boldsymbol{\nabla}_{n}\cdot\left(  \left\vert \psi\right\vert ^{2}%
\frac{\boldsymbol{\nabla}_{n}S}{m_{n}}\right)  =0\ , \label{Cont_psi2_Npart}%
\end{equation}
where $\psi=\left\vert \psi\right\vert e^{iS}$. We can then identify the de
Broglie velocity%
\begin{equation}
\frac{d\mathbf{x}_{n}}{dt}=\mathbf{v}_{n}=\frac{1}{m_{n}}\boldsymbol{\nabla
}_{n}S \label{deB_Npart}%
\end{equation}
of the $n$th particle. Equations (\ref{Sch_Npart}) and (\ref{deB_Npart})
define a deterministic dynamics of a many-body system. If the $i$th and $j$th
particles are entangled, the time evolution of $\mathbf{x}_{i}$ generally
depends instantaneously on $\mathbf{x}_{j}$. The dynamics is nonlocal.

Taking the time derivative of (\ref{deB_Npart}) and using (\ref{Sch_Npart}) we
can write%
\begin{equation}
m_{n}\frac{d^{2}\mathbf{x}_{n}}{dt^{2}}=-\boldsymbol{\nabla}_{n}(V+Q)\ ,
\label{Bohm_acceln}%
\end{equation}
with a `quantum potential'%
\begin{equation}
Q=-\sum_{n=1}^{N}\frac{1}{2m_{n}}\frac{\nabla_{n}^{2}\left\vert \psi
\right\vert }{\left\vert \psi\right\vert }\,\,. \label{Bohm_Q}%
\end{equation}
Bohm regarded (\ref{Bohm_acceln}) as the equation of motion, with in-principle
arbitrary initial conditions for position and momentum. In contrast, in de
Broglie's original dynamics, (\ref{deB_Npart}) is the equation of motion, with
in-principle arbitrary initial conditions for position only. Bohm's dynamics
has been shown to be unstable and not physically tenable (Colin and Valentini
2014). Even so the form (\ref{Bohm_acceln}) is sometimes useful when
discussing the classical limit, which in simple cases can be characterised by
$Q\approx0$ (Holland 1993).

We can now consider an ensemble of similar many-body systems with the same
wave function $\psi$. An arbitrary distribution $\rho(\mathbf{x}%
_{1},\mathbf{x}_{2},...,\mathbf{x}_{N},t)$ satisfies the continuity equation%
\begin{equation}
\frac{\partial\rho}{\partial t}+\sum_{n=1}^{N}\boldsymbol{\nabla}_{n}%
\cdot\left(  \rho\mathbf{v}_{n}\right)  =0\,
\end{equation}
(cf. (\ref{Cont_rho_gen}) with $q=(\mathbf{x}_{1},\mathbf{x}_{2}%
,...,\mathbf{x}_{N})$ and $\partial_{q}=(\boldsymbol{\nabla}_{1}%
,\boldsymbol{\nabla}_{2},...,\boldsymbol{\nabla}_{N})$). This takes the same
form as (\ref{Cont_psi2_Npart}). Thus, as in the general case, if
$\rho=\left\vert \psi\right\vert ^{2}$ initially then $\rho=\left\vert
\psi\right\vert ^{2}$ at later times. This is the state of quantum equilibrium
with the Born rule. Again, in principle there can exist initial quantum
nonequilibrium ensembles with $\rho\neq\left\vert \psi\right\vert ^{2}$.

The theory is easily generalised to a system of $N$ (distinguishable) spin-1/2
particles with a multi-component wave function $\psi_{s_{1}s_{2}...s_{N}%
}(\mathbf{x}_{1},\mathbf{x}_{2},...,\mathbf{x}_{N},t)$ (with spin indices
$s_{1},s_{2},...,s_{N}=+,-$). We have the Schr\"{o}dinger equation (or Pauli
equation)%
\begin{equation}
i\frac{\partial\psi}{\partial t}=\sum_{n=1}^{N}\left(  -\frac{1}{2m_{n}%
}(\boldsymbol{\upsigma}_{n}\cdot\boldsymbol{\nabla}_{n})^{2}+V\right)
\psi\label{Pauli_Nbody}%
\end{equation}
(suppressing spin indices), where the vector Pauli spin matrix
$\boldsymbol{\upsigma}_{n}=(\sigma_{x},\sigma_{y},\sigma_{z})_{n}$ acts on the
$n$th spin index of $\psi$. The wave function generates particle trajectories
$\mathbf{x}_{n}(t)$. The de Broglie velocities $\mathbf{v}_{n}$ follow as
usual from the quantum continuity equation. From (\ref{Pauli_Nbody}) we find a
quantum density $\psi^{\dag}\psi$ and velocities%
\begin{equation}
\mathbf{v}_{n}=\frac{1}{m_{n}}\frac{1}{\psi^{\dag}\psi}\operatorname{Im}%
\psi^{\dag}\boldsymbol{\nabla}_{n}\psi\label{deBNspin}%
\end{equation}
(where $\psi^{\dag}\psi$ is shorthand for $\sum_{s_{1}s_{2}...s_{N}}%
\psi_{s_{1}s_{2}...s_{N}}^{\ast}\psi_{s_{1}s_{2}...s_{N}}$). For a general
ensemble with distribution $\rho(\mathbf{x}_{1},\mathbf{x}_{2},...,\mathbf{x}%
_{N},t)$ we have%
\begin{equation}
\frac{\partial\rho}{\partial t}+\sum_{n=1}^{N}\boldsymbol{\nabla}_{n}%
\cdot(\rho\mathbf{v}_{n})=0\ .
\end{equation}
In this model the particles are not spinning but simply follow trajectories in
space (Bell 1966; 1987, chaps. 1, 15 and 17). For example, for a single
particle in an external magnetic field $\mathbf{B}$, the Hamiltonian includes
a term $\mu\boldsymbol{\upsigma}\cdot\mathbf{B}$, where $\mu$ is the~magnetic
moment, and the gauge-invariant de Broglie velocity includes a term
$-(e/m)\mathbf{A}$.\footnote{For pilot-wave theory in external electromagnetic
fields see Holland (1993).} If the particle enters a Stern-Gerlach apparatus,
it will be deflected upwards or downwards, depending on its initial position,
corresponding to `spin up' and `spin down'.

For $N$ identical particles we have the symmetry condition%
\begin{equation}
\psi_{s_{1}...s_{i}...s_{j}...s_{N}}(\mathbf{x}_{1},...,\mathbf{x}%
_{i},...,\mathbf{x}_{j},...,\mathbf{x}_{N},t)=\pm\psi_{s_{1}...s_{j}%
...s_{i}...s_{N}}(\mathbf{x}_{1},...,\mathbf{x}_{j},...,\mathbf{x}%
_{i},...,\mathbf{x}_{N},t)\ ,
\end{equation}
for distinct $i$, $j=1,...,N$, with $+$ for bosons\ and $-$ for fermions. This
ensures that the set of 3-space velocities is invariant under exchange
$i\leftrightarrow j$ of labels (Bacciagaluppi 2003; Sebens 2016).

\section{Quantum `measurements'}

We consider a so-called quantum `measurement' for a general observable
$\hat{\omega}$. The system (with configuration $q$) is coupled to an apparatus
pointer $y$ via the interaction Hamiltonian%
\begin{equation}
\hat{H}_{\mathrm{I}}=a\hat{\omega}\hat{p}_{y}\ , \label{intn Ham}%
\end{equation}
where $a$ is the coupling and $\hat{p}_{y}=-i\partial_{y}$. During the
interaction $a$ is so large we can neglect the rest of the Hamiltonian. The
initial system wave function $\psi_{0}(q)=\sum_{n}c_{n}\phi_{n}(q)$ is a
superposition of eigenstates ($\hat{\omega}\phi_{n}=\omega_{n}\phi_{n}$), and
the initial pointer packet $g_{0}(y)$ is narrowly-peaked around $y=0$ with
width $\Delta y$. The initial joint wave function $\Psi_{0}(q,y)=\psi
_{0}(q)g_{0}(y)$ evolves to%
\begin{equation}
\Psi(q,y,t)=\sum_{n}c_{n}\phi_{n}(q)g_{0}(y-a\omega_{n}t)\ . \label{Psi-t}%
\end{equation}
The $n$th term shifts by $a\omega_{n}t$ in $y$-space. At sufficiently large
times the `branches' have negligible overlap.

In standard quantum mechanics it is usual to postulate a random `collapse' to
the $n$th branch with probability $p_{n}=\left\vert c_{n}\right\vert ^{2}$. No
such postulate is needed in pilot-wave theory. The configuration
$Q(t)=(q(t),y(t))$ can occupy (the support of) only one of the non-overlapping
terms in (\ref{Psi-t}). Inspection of the de Broglie equations of motion for
$Q$ shows that, after the branches have separated, the velocity $\dot{Q}$ is
determined by the occupied $n$th branch only. Furthermore, because each term
in (\ref{Psi-t}) separates in $q$ and $y$, the system velocity $\dot{q}$ is
determined by $\phi_{n}(q)$ only. At the end of the experiment the system has
an effective `collapsed' wave function%
\begin{equation}
\psi_{\mathrm{coll}}=\phi_{n}(q)\ .
\end{equation}
The probability for this to occur is just the probability $p_{n}$ for $Q(t)$
to occupy the $n$th branch. For an equilibrium ensemble, with distribution
$P(Q,t)=\left\vert \Psi(Q,t)\right\vert ^{2}$, this is found by integrating
$\left\vert \Psi(Q,t)\right\vert ^{2}$ over the support of the $n$th term,
yielding $p_{n}=\left\vert c_{n}\right\vert ^{2}$. For a nonequilibrium
ensemble, with $P(Q,t)\neq\left\vert \Psi(Q,t)\right\vert ^{2}$, in general
$p_{n}\neq\left\vert c_{n}\right\vert ^{2}$. Thus in equilibrium we obtain the
usual Born rule for `measurement' outcomes, while in nonequilibrium the Born
rule is violated.

After the described experiment, there are still empty branches of $\Psi$ in
other regions of configuration space. But these have no effect on the system
for as long as they remain separated from the occupied branch. Subsequent
interaction between $y$ and the environment ensures that a later reoverlap
(with respect to all relevant degrees of freedom) is unlikely (Bohm 1952b, p.
182). For most practical purposes, the `collapse' is irreversible.\footnote{In
principle the empty branches can play a role if a reoverlap somehow occurs.
The scenario of `Wigner's friend' and more recent extensions thereof can be
understood only if we take into account the empty branches (Bacciagaluppi and
Valentini 2009, p. 147; Lazarovici and Hubert 2019).}

A so-called `measurement of $\hat{\omega}$' reduces to an approximate position
measurement for $y$ (to indicate the occupied branch). It is important to
recognise that the `outcome' $\omega_{n}$ is not usually equal to the value of
any previously-existing property of the system, and therefore the quantum
`measurement' is not usually a correct measurement. For example, for a
particle in one dimension with initial wave function $\psi_{0}(x)\propto
e^{ipx}+e^{-ipx}$, a so-called `measurement of momentum' ($\hat{\omega}%
=\hat{p}$) yields outcomes $\pm p$, while the actual initial momentum
$p_{0}=\partial S_{0}/\partial x=0$. Similarly, for the same initial state, a
`measurement of kinetic energy' ($\hat{\omega}=\hat{p}^{2}/2m$) yields a value
$p^{2}/2m$ while the actual value is again $0$ (in this case both before and
after the interaction). It might be said that these measurements are
`unfaithful', which is to say they are not correct measurements. Notable
exceptions are quantum measurements of position ($\hat{\omega}=\hat{x}$),
which do correctly indicate the (previously-existing)\ particle position.

In general quantum `measurements' are not correct measurements. They are
simply experiments that bring about a final state where the system has an
effective wave function $\phi_{n}$. Even so, in quantum equilibrium, there is
no observable difference from standard quantum mechanics: the distribution of
pointer positions is the same. It is only the interpretation that differs.

\section{Quantum relaxation}

We can explain the emergence of the Born rule by a dynamical process of
quantum relaxation. Consider an ensemble of systems with wave function
$\psi(q,t)$, whose configurations $q$ have a distribution $\rho(q,t)$. Because
$\rho$ and $\left\vert \psi\right\vert ^{2}$ obey the same continuity
equation, the ratio $f=\rho/\left\vert \psi\right\vert ^{2}$ obeys an analogue
of the classical Liouville theorem,%
\begin{equation}
\frac{df}{dt}=0\,, \label{f_const}%
\end{equation}
where $d/dt=\partial/\partial t+v\cdot\partial_{q}$ is the time derivative
along trajectories with de Broglie velocity $v$. Despite this, quantum
relaxation can occur for coarse-grained densities $\bar{\rho}$ and
$\overline{\left\vert \psi\right\vert ^{2}}$ (averaged over small
coarse-graining cells),%
\begin{equation}
\bar{\rho}(q,t)\rightarrow\overline{\left\vert \psi(q,t)\right\vert ^{2}}\,\ .
\label{qurel}%
\end{equation}
This can be quantified by a decrease of the coarse-grained $H$-function%
\begin{equation}
\bar{H}(t)=\int dq\ \bar{\rho}\ln(\bar{\rho}/\overline{|\psi|^{2}})
\label{Hqu_cg}%
\end{equation}
(minus the relative entropy of $\bar{\rho}$ with respect to $\overline
{|\psi|^{2}}$), where $\bar{H}\geq0$ and $\bar{H}=0$ if and only if $\bar
{\rho}=\overline{|\psi|^{2}}$. This satisfies a coarse-graining $H$-theorem
(Valentini 1991a)%
\begin{equation}
\bar{H}(t)\leq\bar{H}(0)\ , \label{subquHthm}%
\end{equation}
assuming there is no initial fine-grained structure,%
\begin{equation}
\bar{\rho}(q,0)=\rho(q,0)\ ,\ \ \ \overline{|\psi(q,0)|^{2}}=|\psi
(q,0)|^{2}\ . \label{no micro qu}%
\end{equation}
In fact, $\bar{H}(t)$ strictly decreases ($\bar{H}(t)<\bar{H}(0)$) when $\rho$
develops fine-grained structure ($\rho\neq\bar{\rho}$), which generally occurs
when $v$ varies over the coarse-graining cells (Valentini 1992).

For an ensemble undergoing efficient quantum relaxation, $\bar{H}%
(t)\rightarrow0$ and $\bar{\rho}\rightarrow\overline{|\psi|^{2}}$ (at least to
a good approximation). To show that this actually occurs requires numerical
simulations (Valentini and Westman 2005). The relaxation timescale depends on
the system as well as on the initial conditions.

An example is shown in Fig. 1, for a two-dimensional oscillator whose wave
function is a superposition of 25 energy states (Abraham, Colin, and Valentini
2014). Plotting $\bar{H}$ against $t$ (Fig. 2) we find an approximately
exponential decay,%
\begin{equation}
\bar{H}(t)\approx\bar{H}_{0}e^{-t/\tau}\ , \label{exp decay}%
\end{equation}
with a best-fit timescale $\tau\simeq\allowbreak6$ (in our units), an order of
magnitude larger than the quantum timescale $\Delta t=\hbar/\Delta E\simeq0.5$.%

\begin{figure}
[ptb]
\begin{center}
\includegraphics[
height=2.8237in,
width=4.2203in
]%
{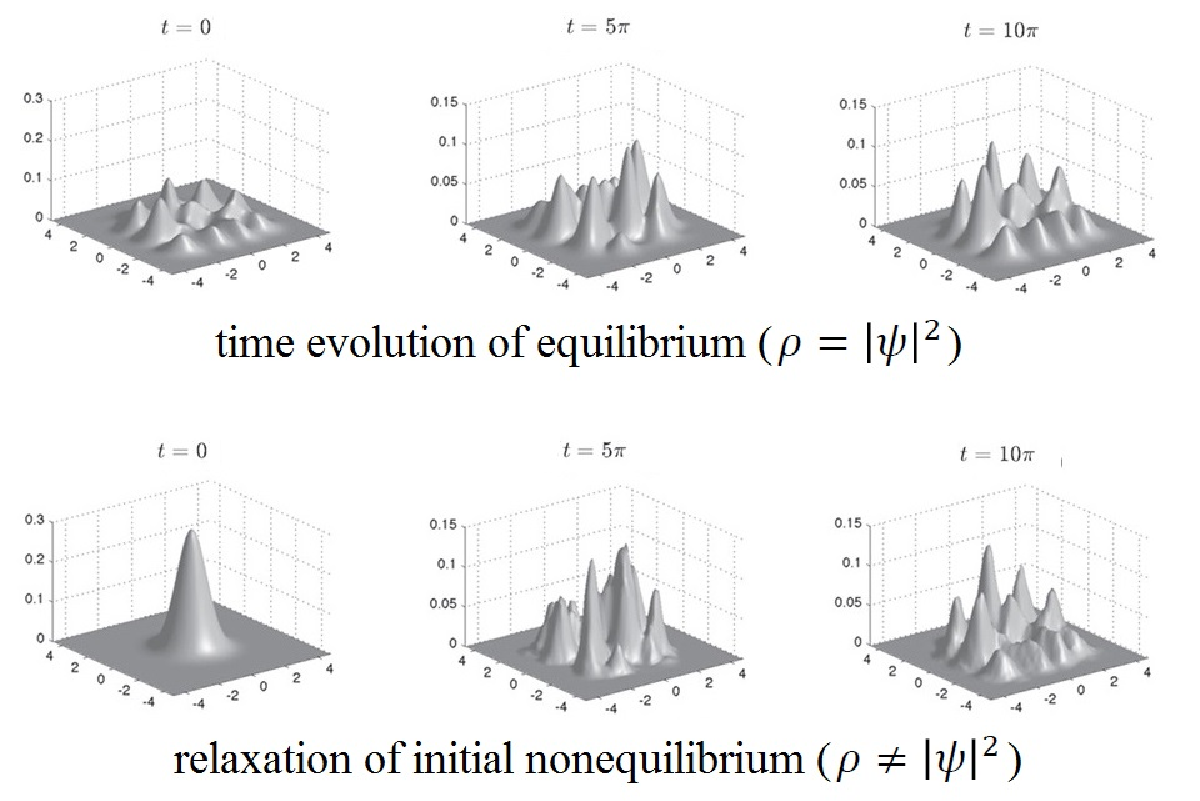}%
\caption{Quantum relaxation for a two-dimensional oscillator (Abraham, Colin,
and Valentini 2014). The initial density $\rho(x,y,0)$ is a Gaussian. The wave
function $\psi(x,y,t)$ has time period $2\pi$ (units $\hbar=m=\omega=1$).}%
\end{center}
\end{figure}

The timescale $\tau$ generally depends on the number $M$ of energy modes in
$\psi$ as well as on the coarse-graining length $\varepsilon$, tending to be
smaller for larger $M$ and for larger $\varepsilon$. For particles in a
two-dimensional box, there is strong evidence for an approximate scaling
$\tau\propto1/M$ (at fixed $\varepsilon$) and weak evidence for $\tau
\sim\varepsilon^{-1/4}$ (at fixed $M$) (Towler, Russell, and Valentini 2012).
Efficient relaxation has been found for $M$ as low as $M=4$ (Abraham, Colin,
and Valentini 2014).

In Fig. 2 the best-fit `residue' $c\simeq0.02$ is comparable to the numerical
error in $\bar{H}$, so the results are consistent with $\bar{H}(t)\rightarrow
0$ as $t\rightarrow\infty$. A small but non-zero residue $\bar{H}%
(t)\rightarrow\bar{H}_{\mathrm{res}}>0$ has been demonstrated for a case with
$M=4$. Residues can occur when the trajectories show significant confinement
(not fully exploring the support of $\left\vert \psi\right\vert ^{2}$).
Whether or not there is significant confinement depends on the initial phases
in $\psi$, and is less likely to occur when $M$ is larger (Abraham, Colin, and
Valentini 2014). Laboratory systems have a long and violent astrophysical
history. In the past $M$ will have been very large, and so we can expect
equilibrium to have been reached to very high accuracy.%

\begin{figure}
[ptb]
\begin{center}
\includegraphics[
height=2.3883in,
width=4.8291in
]%
{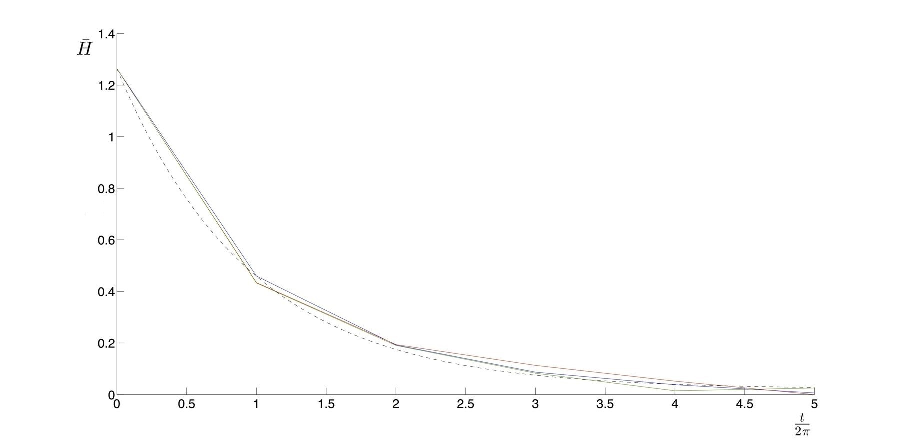}%
\caption{Approximate exponential decay of $\bar{H}(t)$ for a two-dimensional
oscillator (Abraham, Colin, and Valentini 2014). The error is estimated by
running simulations with different grids (solid curves). The dashed curve
shows a best fit to an exponential function $a\exp[-b(t/2\pi)]+c$.}%
\end{center}
\end{figure}

Simulations of quantum relaxation have also been carried out for Dirac
fermions (Colin 2012), for initial extreme nonequilibrium (Underwood 2018),
for systems with small perturbations (Kandhadai and Valentini 2019), and for
coupled systems (Lustosa, Colin, and Perez Bergliaffa 2021; Lustosa,
Pinto-Neto, and Valentini 2023). In general, for relaxing systems the
trajectories tend to be quite erratic, and neighbouring trajectories quickly
diverge. Chaos plays a role in quantum relaxation (Efthymiopoulos,
Contopoulos, and Tzemos 2017; Drezet 2021).

Some authors derive the Born rule seen in the laboratory today from the Born
rule applied to the beginning of the universe (D\"{u}rr, Goldstein, and
Zangh\`{\i} 1992; D\"{u}rr and Teufel 2009). However, this approach is
conceptually circular: the Born rule is assumed at $t=0$ in order to derive
the Born rule at later times (Valentini 1996, 2020). It aids clarity to bear
in mind that, in a deterministic theory of individual systems, initial
conditions are contingencies not fixed by any law or principle. In pilot-wave
theory, whether the universe began in equilibrium or nonequilibrium is
ultimately an empirical question (Section 7.2).

\section{Beyond quantum mechanics}

According to pilot-wave theory the Born rule can be broken, resulting in new
physics outside the domain of standard quantum theory.

\subsection{Nonlocality and nonequilibrium signalling}

Pilot-wave dynamics is nonlocal. Consider for example a pair of entangled
spin-1/2 particles $A$ and $B$. If particle $A$ enters a Stern-Gerlach
apparatus with orientation $\mathbf{m}_{A}$, while particle $B$ enters a
second Stern-Gerlach apparatus with orientation $\mathbf{m}_{B}$, then the
(upwards or downwards) motion of particle $A$ depends nonlocally on the
`measurement setting' $\mathbf{m}_{B}$ of the second apparatus (Bell 1966,
1987). Such nonlocal effects are generic for entangled systems, and are
required by Bell's theorem. Changing the local Hamiltonian $\hat{H}_{B}$ at
one wing of an entangled state instantaneously affects the particle motion
$\mathbf{x}_{A}(t)$ at the other wing, no matter how widely separated the
particles may be.

Now consider an ensemble of entangled pairs in the singlet state. For an
equilibrium ensemble with $\rho(\mathbf{x}_{A},\mathbf{x}_{B},t)=\sum
_{s_{1}s_{2}}\left\vert \psi_{s_{1}s_{2}}(\mathbf{x}_{A},\mathbf{x}%
_{B},t)\right\vert ^{2}$, and for given measurement settings $\mathbf{m}_{A}$,
$\mathbf{m}_{B}$, we find an even ratio of `spin' outcomes $+$, $-$ at $A$ (as
in quantum mechanics). Under a change $\mathbf{m}_{B}\rightarrow\mathbf{m}%
_{B}^{\prime}$ of setting at $B$, some of the outcomes at $A$ will change. A
fraction $\nu_{A}(+,-)$ of the ensemble makes the transition $+\rightarrow-$
at $A$, while a fraction $\nu_{A}(-,+)$ makes the reverse transition
$-\rightarrow+$ at $A$.\footnote{The actual values of the fractions depend on
$\mathbf{m}_{A}$, $\mathbf{m}_{B}$, $\mathbf{m}_{B}^{\prime}$ and have been
calculated in a simplified model of spin (Valentini 2002a).} In equilibrium we
have `detailed balancing':%
\begin{equation}
\nu_{A}(+,-)=\nu_{A}(-,+)\ .
\end{equation}
Despite the change in some individual outcomes at $A$, the marginal
distribution at $A$ remains the same. This is emergent statistical locality:
in equilibrium the nonlocal effects average to zero. In contrast, for a
nonequilibrium ensemble with $\rho(\mathbf{x}_{A},\mathbf{x}_{B},t)\neq
\sum_{s_{1}s_{2}}\left\vert \psi_{s_{1}s_{2}}(\mathbf{x}_{A},\mathbf{x}%
_{B},t)\right\vert ^{2}$, in general%
\begin{equation}
\nu_{A}(+,-)\neq\nu_{A}(-,+)\ ,
\end{equation}
in which case the marginal distribution at $A$ changes and there is a net
statistical nonlocal signal from $B$ to $A$. In nonequilibrium, the nonlocal
effects need not average to zero and statistical locality can be violated
(Valentini 1991b, 2002a). Similar arguments apply to any nonlocal
deterministic hidden-variables theory (Valentini 2002b).

\subsection{Breaking uncertainty}

The uncertainty principle is also a peculiarity of equilibrium and is
generally violated for nonequilibrium ensembles. For example, for a single
particle in one dimension, the outcome $p$ of a quantum `measurement' of
momentum generally depends on the initial particle position $x(0)$ within the
support of the initial wave function $\psi(x,0)$. The distribution of outcomes
$p$ then depends on the distribution $\rho(x,0)$ of initial positions. For
$\rho(x,0)=\left\vert \psi(x,0)\right\vert ^{2}$ the statistical spread
$\Delta p$ matches the prediction of quantum mechanics and is therefore
consistent with the usual uncertainty lower bound $\Delta x\Delta p\geq
\hbar/2$. For $\rho(x,0)\neq\left\vert \psi(x,0)\right\vert ^{2}$ the spread
$\Delta p$ will be anomalous and the uncertainty principle can be violated
(Valentini 1991b, 2025).

\subsection{Subquantum measurement}

Nonequilibrium systems can be employed to perform `subquantum measurements'
whereby we measure the position $x$ of a particle without disturbing its wave
function $\psi_{0}(x)$ (Valentini 2002c). For simplicity we work in one
dimension. We assume the particle has an initial equilibrium probability
density $\rho_{0}(x)=\left\vert \psi_{0}(x)\right\vert ^{2}$. A standard
quantum measurement of $\hat{x}$ will reduce the (effective) wave function to
a narrow packet centred on the measured value. This can be avoided if the
initial pointer coordinate $y$ has a narrow nonequilibrium probability density
$\pi_{0}(y)\neq\left\vert g_{0}(y)\right\vert ^{2}$. If $\pi_{0}(y)$ is
arbitrarily narrow, the initial position $x_{0}$ can be measured without
disturbing $\psi_{0}(x)$, to arbitrary accuracy.

This is done by switching on the interaction Hamiltonian (\ref{intn Ham}),
with $\hat{\omega}=\hat{x}$, for a short time only. During that time we have a
Schr\"{o}dinger equation $\partial\Psi/\partial t=-ax\partial\Psi/\partial y$,
so the initial wave function $\Psi_{0}(x,y)=\psi_{0}(x)g_{0}(y)$ evolves into%
\begin{equation}
\Psi(x,y,t)=\psi_{0}(x)g_{0}(y-axt)\ .
\end{equation}
The continuity equation $\partial\left\vert \Psi\right\vert ^{2}/\partial
t+ax\partial\left\vert \Psi\right\vert ^{2}/\partial y=0$ implies the simple
de Broglie velocities $\dot{x}=0$, $\dot{y}=ax$ and trajectories%
\begin{equation}
x(t)=x_{0},\;y(t)=y_{0}+ax_{0}t\ .
\end{equation}
Now, for $at\rightarrow0$ (with $a$ fixed), $\Psi(x,y,t)\rightarrow\psi
_{0}(x)g_{0}(y)$ and the particle wave function $\psi_{0}(x)$ is undisturbed.
Even so, for arbitrarily small $at$, the pointer $y(t)$ contains precise
information about $x_{0}$, which will be visible to the experimenter if
$\pi_{0}(y)$ is sufficiently narrow. From the continuity equation $\partial
P/\partial t+ax\partial P/\partial y=0$, the initial distribution
$P_{0}(x,y)=\left\vert \psi_{0}(x)\right\vert ^{2}\pi_{0}(y)$ evolves into%
\begin{equation}
P(x,y,t)=\left\vert \psi_{0}(x)\right\vert ^{2}\pi_{0}(y-axt)\ .
\end{equation}
Let $\pi_{0}(y)$ be highly localised around $y=0$, with $\pi_{0}(y)=0$ for
$\left\vert y\right\vert >w/2$. From a standard (faithful) measurement of $y$
we may deduce that $x$ lies in the interval $(y/at-w/2at,\;y/at+w/2at)$ (where
$P(x,y,t)\neq0$ only if $\left\vert y-axt\right\vert <w/2$). Taking
simultaneous limits $at\rightarrow0$ and $w\rightarrow0$, with
$w/at\rightarrow0$, the midpoint $y/at\rightarrow x_{0}$ (since $y=y_{0}%
+ax_{0}t$ and $\left\vert y_{0}\right\vert <w/2$), while the error
$w/2at\rightarrow0$. We then have an exact measurement of $x_{0}$, with no
disturbance of $\psi_{0}(x)$.

An arbitrarily fine time sequence of such measurements can also track the
trajectory $x(t)$ without disturbing the evolving wave function $\psi(x,t)$.

\subsection{Subquantum information and subquantum computation}

In quantum mechanics non-orthogonal quantum states $\left\vert \psi
_{1}\right\rangle $, $\left\vert \psi_{2}\right\rangle $ ($\left\langle
\psi_{1}\right\vert \psi_{2}\rangle\neq0$) cannot be reliably distinguished
for a single system (Nielsen and Chuang 2000). This theorem breaks down in
nonequilibrium pilot-wave theory (Valentini 2002c). The associated sets of
trajectories $q_{1}(t)$, $q_{2}(t)$ generally differ, hence subquantum
monitoring of trajectories can render non-orthogonal states distinguishable.

The E91 quantum-cryptographic protocol (Ekert 1991)\ is rendered insecure if
subquantum measurements allow an eavesdropper to predict quantum `measurement'
outcomes at each wing\ of a (bipartite) entangled state that is being employed
to generate a shared secret key. The B92 protocol (Bennett 1992)\ is rendered
insecure if subquantum measurements allow an eavesdropper to distinguish
non-orthogonal states transmitted between two parties. Thus the discovery of
quantum nonequilibrium systems would undermine the security of the quantum internet.

The ability to distinguish non-orthogonal states reliably also allows new and
potentially more powerful forms of computation. These could be realised in
nonlinear quantum mechanics (Abrams and Lloyd 1998) or in nonequilibrium
pilot-wave theory (Valentini 2002c). However, to evaluate the power of
`subquantum computing' we need to quantify how the nonequilibrium resources
scale (polynomially or exponentially) with the size of the computational task.
Such studies have yet to be carried out.

\section{Quantum field theory and high-energy physics}

We now consider pilot-wave field theory and high-energy physics on flat
spacetime. This is formulated with a preferred rest frame and a preferred time
parameter $t$. Nonlocality acts instantaneously along hypersurfaces of
constant $t$. Lorentz invariance emerges as an effective symmetry for
equilibrium ensembles.

\subsection{Scalar field}

Employing units $c=1$ we first consider a real massive scalar field (Bohm,
Hiley, and Kaloyerou 1987; Valentini 1992; Holland 1993). This has a classical
Lagrangian $L=\int\mathcal{L}d^{3}x$ with%
\begin{equation}
\mathcal{L}={\frac{1}{2}}(\dot{\phi}^{2}-(\boldsymbol{\nabla}\phi)^{2}%
-m^{2}\phi^{2})\ .
\end{equation}
The system has a configuration $q=\phi(\mathbf{x})$. In the functional
Schr\"{o}dinger picture, the wave functional $\Psi\lbrack\phi,t]=\langle
\phi|\Psi(t)\rangle$ (where $|\phi\rangle$ is a field eigenstate) satisfies
the Schr\"{o}dinger equation\footnote{The functional derivative $\delta
\Psi/\delta\phi(\mathbf{x})$ is defined by $\delta\Psi=\int d^{3}%
\mathbf{x}\ \left[  \delta\Psi/\delta\phi(\mathbf{x})\right]  \delta
\phi(\mathbf{x})$ for arbitrary infinitesimal $\delta\phi(\mathbf{x})$.}%
\begin{equation}
i\frac{\partial\Psi}{\partial t}=\int d^{3}x\ \frac{1}{2}\left(  -\frac
{\delta^{2}}{\delta\phi^{2}}+(\boldsymbol{\nabla}\phi)^{2}+m^{2}\phi
^{2}\right)  \Psi\ . \label{Sch_phi}%
\end{equation}
This implies a continuity equation%
\begin{equation}
\frac{\partial|\Psi|^{2}}{\partial t}+\int d^{3}x\ \frac{\delta}{\delta\phi
}\left(  |\Psi|^{2}\frac{\delta S}{\delta\phi}\right)  =0
\end{equation}
(with $\Psi=\left\vert \Psi\right\vert e^{iS}$) and a de Broglie velocity
$v=\delta S/\delta\phi$. We then have the equation of motion%
\begin{equation}
\frac{\partial\phi}{\partial t}=\frac{\delta S}{\delta\phi} \label{deB_phi}%
\end{equation}
for the trajectory $q(t)=\phi(\mathbf{x},t)$. If field elements $\phi
(\mathbf{x})$, $\phi(\mathbf{x}^{\prime})$ at different points $\mathbf{x}%
\neq\mathbf{x}^{\prime}$ are entangled, the time evolution of $\phi
(\mathbf{x})$ generally depends instantaneously on $\phi(\mathbf{x}^{\prime}%
)$. The nonlocal dynamics singles out an absolute simultaneity labelled by the
global time parameter $t$.

For a general ensemble of fields with the same wave functional $\Psi$, the
probability distribution $P[\phi,t]$ evolves by the continuity equation%
\begin{equation}
\frac{\partial P}{\partial t}+\int d^{3}x\ \frac{\delta}{\delta\phi}\left(
P\frac{\delta S}{\delta\phi}\right)  =0\ .
\end{equation}

\subsubsection{Emergent quantum theory and Lorentz invariance}

For an equilibrium ensemble, with $P[\phi,t]=\left\vert \Psi\lbrack
\phi,t]\right\vert ^{2}$, we recover the usual statistical predictions of
quantum field theory. Reference to (\ref{deB_phi}) may be dropped. We may
write (\ref{Sch_phi}) for a quantum state $|\Psi(t)\rangle$ acted on by linear
operators $\hat{\phi}(\mathbf{x})$, $\hat{\pi}(\mathbf{x})$, where $\hat{\phi
}(\mathbf{x})\rightarrow\phi(\mathbf{x})$, $\hat{\pi}(\mathbf{x}%
)\rightarrow-i\delta/\delta\phi(\mathbf{x})$ in the basis $|\phi
(\mathbf{x})\rangle$. Transforming to the Heisenberg picture, operators
$\hat{\omega}(\mathbf{x},t)$ have a time evolution $\partial_{t}\hat{\omega
}=-i[\hat{\omega},\hat{H}]$ with Hamiltonian%
\begin{equation}
\hat{H}=\int d^{3}x\ \frac{1}{2}\left(  \hat{\pi}^{2}+(\boldsymbol{\nabla}%
\hat{\phi})^{2}+m^{2}\hat{\phi}^{2}\right)  \ . \label{Ham_phi}%
\end{equation}
We find $\partial_{t}\hat{\phi}=\hat{\pi}$ and the Lorentz-covariant
Klein-Gordon equation%
\begin{equation}
\partial_{t}^{2}\hat{\phi}-\nabla^{2}\hat{\phi}+m^{2}\hat{\phi}=0\ .
\label{WE1}%
\end{equation}
Lorentz invariance emerges as an equilibrium symmetry, along with statistical
locality and uncertainty.

For a quantum nonequilibrium ensemble, with $P[\phi,t]\neq\left\vert
\Psi\lbrack\phi,t]\right\vert ^{2}$, the usual statistical results of quantum
field theory break down, with potentially observable implications for the
early universe (Section 7). Furthermore, nonlocal signalling is possible and
Lorentz invariance is broken.

At the fundamental level we have an Aristotelian spacetime $E\times E^{3}$
with a preferred state of rest, while in quantum equilibrium we obtain an
effective Minkowski spacetime $M^{4}$. The emergent symmetry group depends,
however, on the structure of the field Hamiltonian. For example if, in
(\ref{Ham_phi}), we replace $(\boldsymbol{\nabla}\hat{\phi})^{2}$ by
$-(\boldsymbol{\nabla}\hat{\phi})^{2}$, we obtain (\ref{WE1}) with a plus sign
in the second term. The symmetries of the operator wave equation then
correspond to an emergent Euclidean spacetime $E^{4}$. As far as we know, for
the Hamiltonians found in nature the emergent spacetime is Lorentzian. We
might ask if there are deeper reasons for this. In some models of particle
physics and quantum gravity, Lorentz invariance breaks down at high energies
(Kosteleck\'{y} and Mewes 2002; Ho\v{r}ava 2009). In a pilot-wave formulation
of such models, Lorentz invariance emerges in equilibrium at low energies only.

\subsubsection{Remarks on Lorentz and Galilean invariance}

Alternatively, some workers attempt to construct a fundamentally
Lorentz-invariant pilot-wave theory. This program has not succeeded. The
dynamics and the quantum equilibrium distribution can be defined consistently
only with a preferred rest frame (Hardy 1992; Berndl et al. 1996). The
definition of the preferred frame might be formally Lorentz covariant
(D\"{u}rr et al. 1999; Tumulka 2007; D\"{u}rr et al. 2014). For example the
preferred frame might coincide with the mean rest frame (Vigier 1985; D\"{u}rr
et al. 2014). While the equations may appear formally Lorentz covariant,
physically there is still a preferred frame.

Some workers regard the low-energy particle theory as Galilean covariant, and
appeal to this purported symmetry to constrain the form of the guidance
equation (\ref{deB_Npart}) (D\"{u}rr, Goldstein, and Zangh\`{\i} 1992).
However, because the dynamics has a law of motion for velocity and not for
acceleration, the natural kinematics is Aristotelian with a preferred state of
rest (Valentini 1997). Galilean covariance is then a fictitious symmetry of
the low-energy theory -- just as covariance under transformation to a
uniformly-accelerated frame is (well known to be) a fictitious symmetry of
Newtonian mechanics. Furthermore, for systems of particles and more generally,
the de Broglie velocity (\ref{deB_gen}) originates from the global symmetry
(\ref{global}) on configuration space and not from any symmetry in space or
spacetime. A natural Aristotelian kinematics is also consistent with a
preferred simultaneity defined by nonequilibrium nonlocal signalling
(Valentini 2008).

\subsubsection{Quantum field relaxation}

The theory of quantum relaxation is readily extended to fields.

It is convenient to work in Fourier space with $\Psi\lbrack\phi_{\mathbf{k}%
},t]$ a function of the Fourier components $\phi_{\mathbf{k}}$. Considering a
box of volume $V$ with periodic boundary conditions, we have a discrete
system. We can write $\phi_{\mathbf{k}}=\frac{\sqrt{V}}{(2\pi)^{3/2}}\left(
q_{\mathbf{k}1}+iq_{\mathbf{k}2}\right)  $, with two real degrees of freedom
$q_{\mathbf{k}r}$ ($r=1$, $2$) for each field mode $\mathbf{k}$. For an
independent (unentangled) field mode, and taking $m=0$, the effective wave
function $\psi_{\mathbf{k}}(q_{\mathbf{k}1},q_{\mathbf{k}2},t)$ satisfies a
two-dimensional Schr\"{o}dinger equation%
\begin{equation}
i\frac{\partial\psi_{\mathbf{k}}}{\partial t}=-\frac{1}{2}\left(
\frac{\partial^{2}}{\partial q_{\mathbf{k}1}^{2}}+\frac{\partial^{2}}{\partial
q_{\mathbf{k}2}^{2}}\right)  \psi_{\mathbf{k}}+\frac{1}{2}k^{2}\left(
q_{\mathbf{k}1}^{2}+q_{\mathbf{k}2}^{2}\right)  \psi_{\mathbf{k}}\ ,
\label{Sch_onemode}%
\end{equation}
while the de Broglie equations of motion for $q_{\mathbf{k}1}$, $q_{\mathbf{k}%
2}$ read%
\begin{equation}
\frac{dq_{\mathbf{k}1}}{dt}=\frac{\partial s_{\mathbf{k}}}{\partial
q_{\mathbf{k}1}},\ \ \ \ \frac{dq_{\mathbf{k}2}}{dt}=\frac{\partial
s_{\mathbf{k}}}{\partial q_{\mathbf{k}2}} \label{deB_onemode}%
\end{equation}
(with $\psi_{\mathbf{k}}=\left\vert \psi_{\mathbf{k}}\right\vert
e^{is_{\mathbf{k}}}$). A general (marginal) distribution $\rho_{\mathbf{k}%
}(q_{\mathbf{k}1},q_{\mathbf{k}2},t)$ obeys%
\begin{equation}
\frac{\partial\rho_{\mathbf{k}}}{\partial t}+\frac{\partial}{\partial
q_{\mathbf{k}1}}\left(  \rho_{\mathbf{k}}\frac{\partial s_{\mathbf{k}}%
}{\partial q_{\mathbf{k}1}}\right)  +\frac{\partial}{\partial q_{\mathbf{k}2}%
}\left(  \rho_{\mathbf{k}}\frac{\partial s_{\mathbf{k}}}{\partial
q_{\mathbf{k}2}}\right)  =0\,. \label{cont_onemode}%
\end{equation}

This system is mathematically equivalent to a two-dimensional oscillator of
mass $m=1$ and angular frequency $\omega=k$. Relaxation results for the
oscillator (Section 4) then apply to a single field mode (Valentini 2007). On
expanding space field relaxation is suppressed at long wavelengths, with
potentially important implications for the early universe (Section 7).

\subsection{Fermions}

Bosons are usually described in terms of fields as above. For fermions, there
are both field-theoretic and particle-trajectory approaches.\footnote{Though
widely seen as problematic, some authors consider particle-trajectory models
for bosons as well.}

\subsubsection{Grassmann field theory}

A pilot-wave theory of a massive spin-1/2 Grassmann field (Valentini 1992,
1996) is conveniently written in terms of a two-component van der Waerden
field, which is equivalent to the more usual four-component Dirac field (Brown
1958; Sakurai 1967).

The two-component complex field $\phi_{\alpha}$ ($\alpha=1,\ 2$) has a
classical Lagrangian $L=\int\mathcal{L}d^{3}x$ with%
\begin{equation}
\mathcal{L}=\dot{\phi}_{\alpha}^{\ast}\dot{\phi}_{\alpha}%
-(\boldsymbol{\upsigma }\cdot\boldsymbol{\nabla}\phi)_{\alpha}^{\ast
}(\boldsymbol{\upsigma}\cdot\boldsymbol{\nabla}\phi)_{\alpha}-m^{2}%
\phi_{\alpha}^{\ast}\phi_{\alpha}\ ,
\end{equation}
where $\sigma^{i}$ are Pauli spin matrices and $\alpha$ is summed over. This
yields the classical wave equation%
\begin{equation}
\partial_{t}^{2}\phi-(\boldsymbol{\upsigma}\cdot\boldsymbol{\nabla})^{2}%
\phi+m^{2}\phi=0
\end{equation}
(suppressing indices). A four-component field $\psi$ obeying the Dirac
equation can be constructed from linear combinations of $\phi$ with its first
space and time derivatives.

In pilot-wave theory $\phi_{\alpha}$, $\phi_{\alpha}^{\ast}$ are taken to be
complex Grassmann fields. These anticommute,%
\begin{equation}
\{\phi_{\alpha}(\mathbf{x}),\ \phi_{\beta}(\mathbf{y})\}=\{\phi_{\alpha}%
^{\ast}(\mathbf{x}),\ \phi_{\beta}^{\ast}(\mathbf{y})\}=\{\phi_{\alpha}^{\ast
}(\mathbf{x}),\ \phi_{\beta}(\mathbf{y})\}=0 \label{eqn52}%
\end{equation}
(where $\{a,\ b\}\equiv ab+ba$), with derivatives defined from the left
($\overrightarrow{\delta}/\delta\phi$) and from the right ($\overleftarrow
{\delta}/\delta\phi$) (Berezin 1966). We have the Schr\"{o}dinger equation for
$\Psi=\Psi\lbrack\phi_{\alpha},\phi_{\alpha}^{\ast},t]$:%
\begin{equation}
i\frac{\partial\Psi}{\partial t}=\int{d^{3}}x{\ }\left(  -\frac
{\overrightarrow{\delta}}{\delta\phi_{\alpha}^{\ast}}\left(  \Psi
\frac{\overleftarrow{\delta}}{\delta\phi_{\alpha}}\right)
+(\boldsymbol{\upsigma}\cdot\boldsymbol{\nabla}\phi)_{\alpha}^{\ast
}(\boldsymbol{\upsigma}\cdot\boldsymbol{\nabla}\phi)_{\alpha}\Psi+m^{2}%
\phi_{\alpha}^{\ast}\phi_{\alpha}\Psi\right)  \ .
\end{equation}
This implies a continuity equation%
\[
\frac{\partial|\Psi|^{2}}{\partial t}+\int{d^{3}x\ }\left(  {\frac
{\overrightarrow{\delta}}{\delta\phi_{\alpha}^{\ast}}}\left(  {|\Psi|^{2}%
\frac{S\overleftarrow{\delta}}{\delta\phi_{\alpha}}}\right)  {+}\left(
{|\Psi|^{2}\frac{\overrightarrow{\delta}S}{\delta\phi_{\alpha}^{\ast}}%
}\right)  {\frac{\overleftarrow{\delta}}{\delta\phi_{\alpha}}}\right)  =0\ ,
\]
from which we can identify the de Broglie velocities%
\begin{equation}
\frac{\partial\phi_{\alpha}^{\ast}}{\partial t}=\frac{S\overleftarrow{\delta}%
}{\delta\phi_{\alpha}}\ ,\ \ \ \ \ \frac{\partial\phi_{\alpha}}{\partial
t}=\frac{\overrightarrow{\delta}S}{\delta\phi_{\alpha}^{\ast}}\ .
\end{equation}

Grassmann variables can be represented by anticommuting matrices. We may then
regard the matrix entries as the evolving objects.

\subsubsection{Dirac sea theory}

An alternative theory of (high-energy) fermions has been developed in terms of
particle trajectories (Bohm, Hiley, and Kaloyerou 1987; Bohm and Hiley 1993;
Colin 2003; Colin and Struyve 2007). In this model there are trajectories for
particles filling the (negative-energy) Dirac sea as well as for particles
with positive energies. Anti-particles are `holes' in the Dirac sea. We
confine ourselves to the free theory. Inclusion of external fields and
interactions is straightforward.

The trajectories are guided by a multi-component wave function $\psi
_{\alpha_{1}\alpha_{2}...\alpha_{n}...}(\mathbf{x}_{1},\mathbf{x}%
_{2},...,\mathbf{x}_{n},...,t)$ (with $\alpha_{n}=1,...,4$) obeying the
many-body Dirac equation%
\begin{equation}
i\frac{\partial\psi}{\partial t}=\sum_{n}\left(  -i\boldsymbol{\upalpha}%
_{n}\cdot\boldsymbol{\nabla}_{n}+\beta_{n}m\right)  \psi
\end{equation}
(suppressing spinor indices), where the $4\times4$ matrices
$\boldsymbol{\upalpha}_{n}$, $\beta_{n}$ act on the $n$th spinor index of
$\psi$. This implies a continuity equation%
\begin{equation}
\frac{\partial(\psi^{\dag}\psi)}{\partial t}+\sum_{n}\boldsymbol{\nabla}%
_{n}\cdot(\psi^{\dag}\boldsymbol{\upalpha}_{n}\psi)=0
\end{equation}
(here $\psi^{\dag}\psi$ is shorthand for $\sum_{\alpha_{1}\alpha_{2}%
...\alpha_{n}...}\psi_{\alpha_{1}\alpha_{2}...\alpha_{n}...}^{\ast}%
\psi_{\alpha_{1}\alpha_{2}...\alpha_{n}...}$). We then have a quantum
equilibrium density $\rho=\psi^{\dag}\psi$ and the de Broglie guidance
equation%
\begin{equation}
\frac{d\mathbf{x}_{n}}{dt}=\frac{\psi^{\dag}\boldsymbol{\upalpha}_{n}\psi
}{\psi^{\dag}\psi}\ .
\end{equation}
More general distributions $\rho\neq\psi^{\dag}\psi$ are in principle
possible. A general density $\rho$ evolves by%
\begin{equation}
\frac{\partial\rho}{\partial t}+\sum_{n}\boldsymbol{\nabla}_{n}\cdot
(\rho\mathbf{v}_{n})=0
\end{equation}
with $\mathbf{v}_{n}=\psi^{\dag}\boldsymbol{\upalpha}_{n}\psi/\psi^{\dag}\psi$.

The above theory of fermions, with deterministic trajectories, was proposed by
Bohm, Hiley, and Kaloyerou (1987). The theory was derived by Colin (2003) as
the continuum limit of a discrete model of fermion numbers evolving
stochastically on a lattice (Bell 1986, 1987). D\"{u}rr et al. (2004, 2005)
studied the same limit of Bell's model but obtained a different result: a
theory of particle trajectories with stochastic jumps at points where
particles are created or annihilated. The difference stems from D\"{u}rr et
al.'s reading of Bell's `fermion number', which was mistakenly taken to be the
number of particles plus the number of anti-particles (instead of Colin's
conventional reading as the number of particles minus the number of anti-particles).

\subsection{Electromagnetic field. Scalar QED}

The pilot-wave theory of the electromagnetic field is often written in the
Coulomb gauge (with a divergence-free vector potential, $\boldsymbol{\nabla
}\cdot\mathbf{A}=0$, and a Coulomb potential $\phi=A^{0}$) (Bohm 1952b;
Kaloyerou 1994). However, on Aristotelian spacetime with a preferred state of
rest, it is more natural to employ a purely spatial gauge field $\mathbf{A}$
with no time component (Valentini 1992, 1996). This is equivalent to working
in the temporal gauge ($A^{0}=0$).\footnote{We can always set $A^{0}=0$ by a
gauge transformation $A^{0}\rightarrow A^{0}+\partial\Lambda/\partial t$,
$\mathbf{A}\rightarrow\mathbf{A}-\boldsymbol{\nabla}\Lambda$ with
$\Lambda(\mathbf{x},t)=-\int_{0}^{t}dt{^{\prime}\ }A^{0}(\mathbf{x},t^{\prime
})$.} Both gauges are non-Lorentz-covariant and hence free of the unphysical
ghost states (some with negative norm) which appear in manifestly
Lorentz-covariant quantum gauge theories (Leibbrandt 1987). In a theory with a
preferred frame, ghosts are automatically avoided.

We can represent the electromagnetic field by a 3-vector field $\mathbf{A}%
(\mathbf{x},t)$ with classical Lagrangian $L=\int\mathcal{L}d^{3}x$ where%
\begin{equation}
\mathcal{L}={\frac{1}{2}}(\mathbf{\dot{A}}^{2}-\mathbf{B}^{2})
\end{equation}
($\mathbf{E}=-\partial_{t}\mathbf{A}$ and $\mathbf{B}=\boldsymbol{\nabla
}\times\mathbf{A}$).{} The field is subject to time-independent gauge
transformations%
\begin{equation}
\mathbf{A}(\mathbf{x},t)\rightarrow\mathbf{A}(\mathbf{x},t)-\boldsymbol{\nabla
}\Lambda(\mathbf{x})\ .
\end{equation}
The wave functional $\Psi=\Psi\lbrack\mathbf{A},t]$ is a function, not on the
space of gauge-dependent fields $\mathbf{A}$, but on the space of equivalence
classes of fields connected by gauge transformations. It then satisfies the
constraint%
\begin{equation}
\boldsymbol{\nabla}\cdot\frac{\delta\Psi}{\delta\mathbf{A}}=0\ , \label{Cons}%
\end{equation}
which follows by setting $\delta\Psi=0$ under an arbitrary infinitesimal gauge
change $\delta\mathbf{A}=-\boldsymbol{\nabla}\Lambda$.

The Schr\"{o}dinger equation is%
\begin{equation}
i\frac{\partial\Psi}{\partial t}=\int d^{3}x\;\frac{1}{2}\left(  -\frac
{\delta^{2}}{\delta\mathbf{A}^{2}}+\mathbf{B}^{2}\right)  \Psi\label{Sch3}%
\end{equation}
(which preserves (\ref{Cons}) in time). This implies a continuity equation%
\begin{equation}
\frac{\partial|\Psi|^{2}}{\partial t}+\int d^{3}x\ \frac{\delta}%
{\delta\mathbf{A}}\cdot\left(  |\Psi|^{2}\frac{\delta S}{\delta\mathbf{A}%
}\right)  =0
\end{equation}
and the guidance equation%
\begin{equation}
\frac{\partial\mathbf{A}}{\partial t}=\frac{\delta S}{\delta\mathbf{A}}\ .
\label{deB3}%
\end{equation}

In quantum equilibrium, again, reference to (\ref{deB3}) may be dropped, and
the field statistics obey the Born rule. We then recover standard QED in the
temporal gauge. In the Heisenberg picture, (\ref{Cons}) becomes%
\begin{equation}
\boldsymbol{\nabla}\cdot\partial_{t}\mathbf{\hat{A}}\left\vert \Psi
\right\rangle =0\ . \label{ConsH}%
\end{equation}
We have equal-time commutation relations\footnote{On Aristotelian spacetime
the spatial indices $i$, $j$ can be trivially raised or lowered with no sign
change.}%
\begin{equation}
\left[  \hat{A}_{i}(\mathbf{x},t),\partial_{t}\hat{A}_{j}(\mathbf{x}^{\prime
},t)\right]  =i\delta_{ij}\delta^{3}(\mathbf{x}-\mathbf{x}^{\prime})
\label{CRs}%
\end{equation}
($i$, $j=1,2,3$) and the operator field equation%
\begin{equation}
\partial_{t}^{2}\mathbf{\hat{A}}+\boldsymbol{\nabla}\times(\boldsymbol{\nabla
}\times\mathbf{\hat{A}})=0\ . \label{WE2}%
\end{equation}
Equations (\ref{ConsH})--(\ref{WE2}) define (free) QED in the temporal gauge,
which is equivalent to the more usual forms of QED (in the Coulomb gauge or a
Lorentz-covariant gauge) (Leibbrandt 1987; Valentini 1996).

Interactions can be introduced in the usual way via local gauge invariance.
For simplicity we consider interaction with a charged scalar field (scalar
QED) (Valentini 1992). The gauge transformations%
\begin{equation}
\phi\rightarrow\phi\mathrm{e}^{ie\Lambda},\;\;\;\phi^{\ast}\rightarrow
\phi^{\ast}\mathrm{e}^{-ie\Lambda},\;\;\;\mathbf{A}\rightarrow\mathbf{A}%
-\boldsymbol{\nabla}\Lambda\label{GS}%
\end{equation}
(with $\Lambda=\Lambda(\mathbf{x})$ and $e$ the charge) imply that $\Psi
=\Psi\lbrack\phi,\phi^{\ast},\mathbf{A},t]$ is subject to the constraint%
\begin{equation}
\boldsymbol{\nabla}\cdot\frac{\delta\Psi}{\delta\mathbf{A}}=ie\left(
\phi^{\ast}\frac{\delta\Psi}{\delta\phi^{\ast}}-\phi\frac{\delta\Psi}%
{\delta\phi}\right)  \label{Cons4}%
\end{equation}
(again obtained by setting $\delta\Psi=0$ under an infinitesimal gauge
change). The Schr\"{o}dinger equation is%
\begin{equation}
i\frac{\partial\Psi}{\partial t}=\int d^{3}\mathbf{x}\;\left(  -\frac{1}%
{2}\frac{\delta^{2}}{\delta\mathbf{A}^{2}}+\frac{1}{2}\mathbf{B}^{2}%
-\frac{\delta^{2}}{\delta\phi^{\ast}\delta\phi}+m^{2}\phi^{\ast}%
\phi+(\mathbf{D}\phi)^{\ast}\cdot(\mathbf{D}\phi)\right)  \Psi\ , \label{Sch4}%
\end{equation}
where $\mathbf{D}\phi=\boldsymbol{\nabla}\phi+ie\mathbf{A}\phi$ is the
gauge-covariant derivative. From the associated continuity equation we find
the guidance equations%
\[
\frac{\partial\phi}{\partial t}=\frac{\delta S}{\delta\phi^{\ast}}%
,\;\;\;\frac{\partial\phi^{\ast}}{\partial t}=\frac{\delta S}{\delta\phi
},\;\;\;\frac{\partial\mathbf{A}}{\partial t}=\frac{\delta S}{\delta
\mathbf{A}}\ .
\]
In quantum equilibrium we recover standard (scalar) QED in the temporal gauge.

To describe interactions with electrons and positrons we can introduce a gauge
coupling of $\mathbf{A}$ to a spin-1/2 complex Grassmann field or (in a
particle description of fermions) to the many-body Dirac equation. Again, in
quantum equilibrium, we necessarily recover standard QED in the temporal gauge.

\subsection{Non-Abelian gauge theories}

In the standard SU(3)$\times$SU(2)$\times$U(1) model of particle physics,
interactions are usually mediated by 4-vector gauge fields on Minkowski
spacetime, as defined in a manifestly Lorentz-covariant quantum field theory.
In pilot-wave theory it is more natural to employ 3-vector gauge fields on
Aristotelian spacetime (Valentini 1992, 1996). In quantum equilibrium we
recover the usual theories of electroweak and strong interactions written in
the ghost-free temporal gauge (with all time components set to zero).

\subsubsection{QCD}

In quantum chromodynamics we have 3-vector gluon fields $\mathbf{A}%
^{a}(\mathbf{x},t)$ ($a=1,...,8$) with a classical Lagrangian $L=\int
\mathcal{L}d^{3}x$ where\footnote{Repeated indices are summed.}%
\begin{equation}
\mathcal{L}={\frac{1}{2}}(\dot{A}_{i}^{a})^{2}-{\frac{1}{4}(}F_{ij}^{a}%
)^{2}\ ,
\end{equation}
with%
\begin{equation}
F_{ij}^{a}=-\partial_{i}A_{j}^{a}+\partial_{j}A_{i}^{a}+gf^{abc}A_{i}^{b}%
A_{j}^{c}\ ,
\end{equation}
where the fields are subject to time-independent (infinitesimal) SU(3) gauge
transformations%
\begin{equation}
\mathbf{A}^{a}\rightarrow\mathbf{A}^{a}-f^{abc}\Lambda^{b}\mathbf{A}^{c}%
-\frac{1}{g}\boldsymbol{\nabla}\Lambda^{a}\ .\label{gauge_QCD}%
\end{equation}
Here $\Lambda^{a}$ are arbitrary (infinitesimal) functions on space, $f^{abc}$
are the structure constants, and $g$ is the gauge coupling. For free gluon
fields the wave functional $\Psi=\Psi\lbrack\mathbf{A}^{a},t]$ is then
constrained by%
\begin{equation}
\boldsymbol{\nabla}\cdot\frac{\delta\Psi}{\delta\mathbf{A}^{a}}=gf^{abc}%
\mathbf{A}^{b}\cdot\frac{\delta\Psi}{\delta\mathbf{A}^{c}}%
\end{equation}
(setting $\delta\Psi=0$ under transformations (\ref{gauge_QCD})).

The Schr\"{o}dinger equation is%
\begin{equation}
i\frac{\partial\Psi}{\partial t}=\int d^{3}x\;\frac{1}{2}\left(  -\frac
{\delta^{2}}{(\delta A_{i}^{a})^{2}}+\frac{1}{2}{(}F_{ij}^{a})^{2}\right)
\Psi\ .
\end{equation}
The associated continuity equation for $\left\vert \Psi\right\vert ^{2}$
implies the de Broglie velocities%
\begin{equation}
\frac{\partial\mathbf{A}^{a}}{\partial t}=\frac{\delta S}{\delta\mathbf{A}%
^{a}}\ .
\end{equation}

To describe interactions with quarks we can introduce an SU(3) gauge coupling
to colour triplets of spin-1/2 Grassmann fields or (in a particle description
of fermions) to colour triplets of the many-body Dirac equation. Again, in
quantum equilibrium, we recover standard QCD in the temporal gauge.

\subsubsection{Electroweak Theory. Spontaneous Symmetry Breaking}

A pilot-wave formulation of $\mathrm{SU}(2)\times\mathrm{U}(1)$ electroweak
theory is also readily obtained from the standard theory in the temporal
gauge. We now have four massless 3-vector gauge fields $\mathbf{W}_{i}$
($i=1,\ 2,\ 3$) and $\mathbf{B}$ (not to be confused with the magnetic field).
These are related to the $W$-boson and $Z^{0}$-boson fields $\mathbf{W}%
,\ \mathbf{W}^{\dag},\ \mathbf{Z}$, and to the electromagnetic field
$\mathbf{A}$, by the usual relations%
\begin{equation}
\mathbf{W}={\frac{1}{\sqrt{2}}}(\mathbf{W}_{1}-i\mathbf{W}_{2}%
)\ ,\ \ \ \ \ \mathbf{W}^{\dag}={\frac{1}{\sqrt{2}}}(\mathbf{W}_{1}%
+i\mathbf{W}_{2})\ ,
\end{equation}%
\begin{equation}
\mathbf{Z}=\mathbf{W}_{3}\cos\theta_{W}-\mathbf{B}\sin\theta_{W}%
\ ,\ \ \ \ \ \mathbf{A}=\mathbf{W}_{3}\sin\theta_{W}+\mathbf{B}\cos\theta
_{W}\ ,
\end{equation}
where $\theta_{W}$ is the weak mixing angle. The fields $\mathbf{W}%
,\ \mathbf{W}^{\dag},\ \mathbf{Z}$ become effectively massive after
spontaneous symmetry breaking. In quantum equilibrium we necessarily recover
standard electroweak theory in the temporal gauge, with time components
$W^{0}=(W^{\dag})^{0}=Z^{0}=A^{0}=0$.

The pilot-wave account of spontaneous symmetry breaking is, again, as in the
standard theory written in the temporal gauge. For example, the Abelian Higgs
model is just scalar QED with the Hamiltonian in (\ref{Sch4}) supplemented by
a self-interaction term $\lambda\left\vert \phi\right\vert ^{4}$. As usually
understood, in appropriate conditions the system has a continuous set of
stable ground states with non-zero (constant) expectation values $\phi_{0}$
for the field $\phi$, where an arbitrary choice of ground state effectively
breaks the gauge symmetry. Expanding $\phi$ around $\phi_{0}$, the term
$(ie\mathbf{A}\phi)^{\ast}\cdot(ie\mathbf{A}\phi)$ in the original Hamiltonian
yields an effective mass for the field $\mathbf{A}$, arising from the coupling
between $\mathbf{A}$ and the constant field $\phi_{0}$. The effective particle
content -- massive neutral vector bosons and massive neutral scalar bosons --
is usually shown in the unitary gauge. But the Higgs model in the temporal
gauge is equivalent to the model in the unitary gauge (Kim et al. 1990). In
quantum equilibrium the pilot-wave formulation, with 3-vector gauge fields, is
then necessarily equivalent to the standard theory.

\section{The early universe}

Quantum field relaxation presumably took place in the early universe. To
describe this we need to generalise pilot-wave field theory to a background
curved spacetime. On expanding space we find that quantum relaxation can be
suppressed at long wavelengths, so that early violations of the Born rule may
be imprinted on the cosmic microwave background (CMB) at large scales.

\subsection{Quantum fields in curved spacetime}

Pilot-wave field theory is readily generalised to a classical curved spacetime
background (Valentini 2004).

We assume spacetime to be globally hyperbolic, so it can be foliated
(nonuniquely) by spacelike hypersurfaces $\Sigma(t)$ with a global time $t$.
The spacetime line element can then be written as%
\begin{equation}
d\tau^{2}=(N^{2}-N_{i}N^{i})dt^{2}-2N_{i}dx^{i}dt-g_{ij}dx^{i}dx^{j}\ ,
\label{ADM}%
\end{equation}
with a lapse function $N$ and shift vector $N^{i}$, where $g_{ij}$ is the
3-metric on $\Sigma(t)$. We may set $N^{i}=0$ (provided lines $x^{i}%
=\mathrm{const}.$ do not meet singularities).

For a massless, minimally-coupled real scalar field $\phi$, we have a
classical Lagrangian $L=\int\mathcal{L}d^{3}x$ with%
\begin{equation}
\mathcal{L}=\frac{1}{2}N\sqrt{g}\left(  \frac{1}{N^{2}}\dot{\phi}^{2}%
-\,g^{ij}\partial_{i}\phi\partial_{j}\phi\right)  \ ,
\end{equation}
where $g=\det g_{ij}$. The wave functional $\Psi\lbrack\phi,t]$ satisfies the
Schr\"{o}dinger equation%
\begin{equation}
i\frac{\partial\Psi}{\partial t}=\int d^{3}x\;\frac{1}{2}N\sqrt{g}\left(
-\frac{1}{g}\frac{\delta^{2}}{\delta\phi^{2}}+g^{ij}\partial_{i}\phi
\partial_{j}\phi\right)  \Psi\ . \label{Sch2}%
\end{equation}
The associated continuity equation for $\left\vert \Psi\right\vert ^{2}$
implies the de Broglie velocity%
\begin{equation}
\frac{\partial\phi}{\partial t}=\frac{N}{\sqrt{g}}\frac{\delta S}{\delta\phi
}\ . \label{deB2}%
\end{equation}
An arbitrary distribution $P[\phi,t]$ satisfies the same continuity equation,%
\begin{equation}
\frac{\partial P}{\partial t}+\int d^{3}x\;\frac{\delta}{\delta\phi}\left(
P\frac{N}{\sqrt{g}}\frac{\delta S}{\delta\phi}\right)  =0\ .
\label{cont_QFICS}%
\end{equation}
As usual, $P=\left\vert \Psi\right\vert ^{2}$ at some initial time implies
$P=\left\vert \Psi\right\vert ^{2}$ at all times.

If $\Psi$ is entangled across the hypersurface $\Sigma(t)$, the time evolution
of $\phi$ at a spatial point $x^{i}$ can depend instantaneously (with respect
to $t$) on $\phi$ at remote points $(x^{\prime})^{i}\neq x^{i}$. For
nonequilibrium ensembles, there will be statistical nonlocal signals across
$\Sigma(t)$. The theory is physically consistent if we assume there is a
preferred foliation (Valentini 2008).

\subsection{Suppression of quantum relaxation on expanding space}

Quantum relaxation in the early universe has been studied for a massless field
$\phi$ on flat expanding space, with spacetime metric%
\begin{equation}
d\tau^{2}=dt^{2}-a^{2}\delta_{ij}dx^{i}dx^{j}\ , \label{exp space}%
\end{equation}
and with a scale factor $a(t)\propto t^{1/2}$ (a radiation-dominated
expansion) (Colin and Valentini 2013, 2015). In Fourier space, for an
unentangled field mode $(q_{\mathbf{k}1},q_{\mathbf{k}2})$, we again obtain
Schr\"{o}dinger and de Broglie equations of the form (\ref{Sch_onemode}) and
(\ref{deB_onemode}) for a two-dimensional oscillator but with mass $m=a^{3}$
and angular frequency $\omega=k/a$ (Valentini 2007). This system is equivalent
to a standard oscillator, of constant mass and angular frequency, with $t$
replaced by a `retarded time' $t_{\mathrm{ret}}=t_{\mathrm{ret}}(t,k)$ (Colin
and Valentini 2013). Quantum relaxation depends on the relation between the
physical wavelength $\lambda_{\mathrm{phys}}=a(2\pi/k)$ and the Hubble radius
$H^{-1}=a/\dot{a}$. For $\lambda_{\mathrm{phys}}<<H^{-1}$ we find
$t_{\mathrm{ret}}(t,k)\rightarrow t$ and the usual rapid relaxation (Fig. 1).
For $\lambda_{\mathrm{phys}}\gtrsim H^{-1}$ we find $t_{\mathrm{ret}}(t,k)<t$
and relaxation is suppressed.

The suppression of quantum relaxation can be quantified by the mean square%
\begin{equation}
\left\langle \left\vert \phi_{\mathbf{k}}\right\vert ^{2}\right\rangle
=\left\langle \left\vert \phi_{\mathbf{k}}\right\vert ^{2}\right\rangle
_{\mathrm{QT}}\xi(k)\ ,
\end{equation}
where $\left\langle ...\right\rangle $ and $\left\langle ...\right\rangle
_{\mathrm{QT}}$ are respectively nonequilibrium and equilibrium expectation
values and $\xi(k)<1$. Numerical simulations show that $\xi(k)$ tends to
decrease as $k$ decreases (ignoring small oscillations), with a good fit to
the curve%
\begin{equation}
\xi(k)=\tan^{-1}(c_{1}\frac{k}{\pi}+c_{2})-\frac{\pi}{2}+c_{3}\ , \label{atan}%
\end{equation}
where the parameters $c_{1}$, $c_{2}$, $c_{3}$ depend on the initial state and
on the time interval (Colin and Valentini 2015, 2016).

In inflationary cosmology, during a very early approximately exponential
expansion, perturbations $\phi_{\mathbf{k}}$ of the inflaton field generate
primordial curvature perturbations $\mathcal{R}_{\mathbf{k}}\propto
\phi_{\mathbf{k}}$, which later generate small temperature anisotropies in the
CMB (Liddle and Lyth 2000). By measuring statistical properties of the CMB we
can constrain the primordial power spectrum%
\begin{equation}
\mathcal{P}_{\mathcal{R}}(k)=\frac{4\pi k^{3}}{V}\left\langle |\mathcal{R}%
_{\mathbf{k}}|^{2}\right\rangle \propto\frac{4\pi k^{3}}{V}\left\langle
|\phi_{\mathbf{k}}|^{2}\right\rangle \ .
\end{equation}
Thus from measurements of the CMB we can set observational bounds on
corrections to the Born rule in the very early universe (Valentini 2010).

Assuming that quantum relaxation occurred during a pre-inflationary era, and
making the simplifying assumption that the spectrum is unaffected by the
transition from pre-inflation to inflation, the usual quantum spectrum
$\mathcal{P}_{\mathcal{R}}^{\mathrm{QT}}(k)$ is replaced by%
\begin{equation}
\mathcal{P}_{\mathcal{R}}(k)=\mathcal{P}_{\mathcal{R}}^{\mathrm{QT}}%
(k)\xi(k)\ ,
\end{equation}
with a power deficit at small $k$ of the form (\ref{atan}). CMB data show
hints of a power deficit at small $k$ (Aghanim et al. 2016). However, current
data neither support nor rule out the prediction (\ref{atan}) (Vitenti, Peter,
and Valentini 2019). Further predictions, such as nonequilibrium violations of
statistical isotropy, are needed to obtain sufficient evidence for or against
the model.

According to inflationary cosmology, the matter in our universe was created at
early times by inflaton decay. If the early inflaton field violates the Born
rule, so will its decay products (Underwood and Valentini 2015). Such early
nonequilibrium particles may have survived to the present day, possibly as a
component of dark matter. If those particles decay or annihilate, the
resulting spectral lines could be anomalous (Underwood and Valentini 2020).
The decay or annihilation products (probably photons) may show violations of
the Born rule in simple quantum experiments.

\section{Quantum gravity and quantum cosmology}

The pilot-wave theory of gravitation, with de Broglie-Bohm trajectories
associated with the time-independent Wheeler-DeWitt equation, was first
discussed for quantum cosmology by Vink (1992) and then developed for quantum
gravity by Horiguchi (1994).\footnote{An alternative approach attempts to
associate trajectories with a time-dependent gravitational Schr\"{o}dinger
equation (Valentini 1992, 1996; Roser and Valentini 2014).} It has since been
extensively applied to cosmology (Pinto-Neto 2005; Pinto-Neto and Fabris 2013;
Pinto-Neto 2021).

\subsection{Pilot-wave theory and the Wheeler-DeWitt equation}

Beginning with the classical Einstein-Hilbert action, canonical quantum
gravity applies the foliation (\ref{ADM}) of spacetime and quantises the
spatial 3-geometry with metric $g_{ij}$. In the presence of a scalar matter
field $\phi$ with potential $\mathcal{V}(\phi)$, this yields the
Wheeler-DeWitt equation (units $\hbar=c=16\pi G=1$) (Kiefer 2012)%
\begin{equation}
\left(  -G_{ijkl}\frac{\delta^{2}}{\delta g_{ij}\delta g_{kl}}-\sqrt
{g}R+\mathcal{\hat{H}}_{\phi}\right)  \Psi=0 \label{WD}%
\end{equation}
for the wave functional $\Psi=\Psi\lbrack g_{ij},\phi]$, where%
\begin{equation}
G_{ijkl}={\frac{1}{2}}g^{-1/2}(g_{ik}g_{jl}+g_{il}g_{jk}-g_{ij}g_{kl})
\end{equation}
and%
\begin{equation}
\mathcal{\hat{H}}_{\phi}=\frac{1}{2}\sqrt{g}\left(  -\frac{1}{g}\frac
{\delta^{2}}{\delta\phi^{2}}+g^{ij}\partial_{i}\phi\partial_{j}\phi\right)
+\sqrt{g}\mathcal{V}\ .
\end{equation}
The ordering in the kinetic term `$G\delta^{2}/\delta g^{2}$' is ambiguous. We
also have a constraint%
\begin{equation}
-2D_{j}\frac{\delta\Psi}{\delta g_{ij}}+\partial^{i}\phi\frac{\delta\Psi
}{\delta\phi}=0 \label{cons_QG}%
\end{equation}
(where $D_{j}$ is a spatial covariant derivative), which follows by setting
$\delta\Psi=0$ under arbitrary infinitesimal spatial diffeomorphisms. This
ensures that $\Psi$ is a function of the coordinate-independent 3-geometry and
not of the coordinate-dependent 3-metric.

Writing (\ref{WD}) with the operator ordering `$(\delta/\delta g)G(\delta
/\delta g)$', inserting $\Psi=\left\vert \Psi\right\vert e^{iS}$ and taking
the imaginary part, we obtain%
\begin{equation}
\frac{\delta}{\delta g_{ij}}\left(  |\Psi|^{2}2NG_{ijkl}\frac{\delta S}{\delta
g_{kl}}\right)  +\frac{\delta}{\delta\phi}\left(  |\Psi|^{2}\frac{N}{\sqrt{g}%
}\frac{\delta S}{\delta\phi}\right)  =0\ . \label{WD_cont}%
\end{equation}
This is not a continuity equation but an infinity of equations (one per space
point $x^{i}$). Even so we may take it to define a natural velocity field in
configuration space, and so identify the de Broglie guidance equations%
\begin{equation}
\frac{\partial g_{ij}}{\partial t}=2NG_{ijkl}\frac{\delta S}{\delta g_{kl}%
}\ ,\ \ \ \ \frac{\partial\phi}{\partial t}=\frac{N}{\sqrt{g}}\frac{\delta
S}{\delta\phi}\ . \label{deB_QG}%
\end{equation}
Alternatively, these can be derived by setting the classical canonical momenta
equal to the phase gradient.\footnote{For $N_{i}\neq0$ the right-hand sides of
(\ref{deB_QG}) have additional terms $D_{i}N_{j}+D_{j}N_{i}$ and
$N^{i}\partial_{i}\phi$ respectively.} Equations (\ref{deB_QG}) and
(\ref{WD}), with the constraint (\ref{cons_QG}), are taken to define the
dynamics of an individual system.

\subsection{Dynamical consistency. Preferred foliation}

The consistency of the above dynamics has been questioned. For given initial
conditions on a spacelike slice, the 4-geometry traced out by the evolving
3-metric should not depend on the arbitrary lapse and shift functions $N$,
$N^{i}$. Otherwise the initial-value problem would not be well posed. Shtanov
(1996) argued that the resulting 4-geometry depends on $N$ and that the theory
breaks foliation invariance. We might then make a case for including a
specific choice for $N$ as part of the theory. In contrast, work by Pinto-Neto
and Santini (2002) appears to demonstrate that the 4-geometry is in fact
independent of $N$, $N^{i}$. By writing the dynamics as a classical
Hamiltonian system (with an additional `quantum potential density'), and
applying the analogue of a well-known theorem from the Hamiltonian dynamics of
classical gravitation, Pinto-Neto and Santini argue that the evolving 3-metric
yields the same 4-geometry for all choices of $N$, $N^{i}$ -- a 4-geometry
that is, however, non-Lorentzian. The argument seems to imply that the
resulting spacetime has an effective preferred foliation (for a given wave
functional $\Psi$, and for a given initial 3-geometry). Local Lorentz
invariance is broken, as we would expect in a nonlocal theory.

These results seem consistent with the first-order (Aristotelian) structure of
pilot-wave dynamics, which naturally defines a preferred state of rest
(Valentini 1997). Furthermore, nonequilibrium entangled systems generate
statistical nonlocal signals (Section 5.1), which may be employed to define an
absolute simultaneity (Valentini 2008). Thus, general features of pilot-wave
theory already point to an underlying preferred foliation of spacetime.

\subsection{Pilot-wave quantum cosmology}

As a simple example consider a flat expanding universe, with line element
(\ref{exp space}), containing a homogenous matter field $\phi$ with potential
$\mathcal{V}(\phi)$. With appropriate factor ordering, the Wheeler-DeWitt
equation for $\Psi(a,\phi)$ reads (Brizuela, Kiefer, and Kr\"{a}mer 2016)%

\begin{equation}
\frac{1}{2m_{\mathrm{P}}^{2}}\frac{1}{a}\frac{\partial}{\partial a}\left(
a\frac{\partial\Psi}{\partial a}\right)  -\frac{1}{2a^{2}}\frac{\partial
^{2}\Psi}{\partial\phi^{2}}+a^{4}\mathcal{V}\Psi=0\ , \label{WD_mini}%
\end{equation}
where $m_{\mathrm{P}}^{2}=3/4\pi G$ is the (rescaled) Planck mass squared.
Inserting $\Psi=\left\vert \Psi\right\vert e^{iS}$ and taking the imaginary
part, we find a continuity equation%
\begin{equation}
-\frac{1}{m_{\mathrm{P}}^{2}}\frac{\partial}{\partial a}\left(  a^{2}%
\left\vert \Psi\right\vert ^{2}\frac{1}{a}\frac{\partial S}{\partial
a}\right)  +\frac{\partial}{\partial\phi}\left(  a^{2}\left\vert
\Psi\right\vert ^{2}\frac{1}{a^{3}}\frac{\partial S}{\partial\phi}\right)  =0
\label{p-cont_qucosmo_t}%
\end{equation}
for a static density $a^{2}\left\vert \Psi\right\vert ^{2}$ with a natural
velocity field%
\begin{equation}
\dot{a}=-\frac{1}{m_{\mathrm{P}}^{2}}\frac{1}{a}\frac{\partial S}{\partial
a}\ ,\ \ \dot{\phi}=\frac{1}{a^{3}}\frac{\partial S}{\partial\phi}\ .
\label{deB_qu_cosm_t}%
\end{equation}
We can identify (\ref{deB_qu_cosm_t}) as the de Broglie guidance equations.
Again, the same equations follow if we set the classical canonical momenta
equal to the phase gradient.

De Broglie-Bohm quantum cosmology has been extensively applied to bouncing
cosmological models (alternatives to inflation with a contracting era followed
by an expanding phase) and to the theory of cosmological perturbations. Other
applications include demonstrations of singularity avoidance, explanations for
cosmic acceleration, and a clear treatment of the quantum-to-classical
transition in the early universe. For reviews see Pinto-Neto (2005),
Pinto-Neto and Fabris (2013), and Pinto-Neto (2021).

\subsection{Problem of probability. Beyond the Born rule}

Most work in pilot-wave quantum cosmology is focussed on properties of
trajectories, without developing a theory of ensembles or probabilities. This
is because $\Psi$ is non-normalisable and cannot define a Born-rule density.
This is a general feature of solutions to the Wheeler-DeWitt equation, whose
mathematical structure is analogous to a Klein-Gordon equation.\footnote{Owing
to the indefinite `DeWitt metric' $G_{ijkl}$.} The integral $\int
dq\ \left\vert \Psi\right\vert ^{2}$ over configuration space is comparable to
the integral $\int d^{3}x\int dt\ \left\vert \phi\right\vert ^{2}$ for a
Klein-Gordon field $\phi$ over spacetime and necessarily diverges (even after
factoring out coordinate dependence) (Isham 1993; Kucha\v{r} 2011). Even so,
we can make sense of probability in pilot-wave quantum gravity (Valentini
2021, 2023).

First, we can define a theoretical ensemble with a normalised probability
density $P$, even though $P$ can never be equal to the (non-normalised)
Born-rule density $\left\vert \Psi\right\vert ^{2}$. For example, in our model
of quantum cosmology, we can write a general continuity equation%
\begin{equation}
\frac{\partial P}{\partial t}+\frac{\partial}{\partial a}\left(  P\dot
{a}\right)  +\frac{\partial}{\partial\phi}\left(  P\dot{\phi}\right)  =0\ ,
\end{equation}
with de Broglie velocities (\ref{deB_qu_cosm_t}). This can be applied to a
general theoretical ensemble with wave function $\Psi$. Because $P$ is by
construction normalised, it can never relax to $\left\vert \Psi\right\vert
^{2}$ (or to $a^{2}\left\vert \Psi\right\vert ^{2}$ in our model). Thus the
deep quantum-gravity regime is always in a state of quantum nonequilibrium.

Second, the Born rule can emerge in a semiclassical regime, with an effective
time-dependent Schr\"{o}dinger equation $i\partial\psi/\partial t=\hat{H}\psi$
for a normalisable wave functional $\psi\lbrack\phi,t]$, where $\phi$
represents quantum fields propagating on a classical spacetime background and
$\hat{H}$ is the effective Hamiltonian. This approximation arises with a
Wheeler-DeWitt wave functional $\Psi\lbrack g_{ij},\phi]\approx\Psi
_{\mathrm{WKB}}[g_{ij}]\psi\lbrack\phi,g_{ij}]$, where $\Psi_{\mathrm{WKB}%
}[g_{ij}]$ is a WKB state and the effective $\psi\lbrack\phi,t]$ is just
$\psi\lbrack\phi,g_{ij}(t)]$ evaluated along a classical background trajectory
$g_{ij}(t)$. In this regime we find the usual (coarse-grained)\ relaxation
$\rho\rightarrow\left\vert \psi\right\vert ^{2}$ to the Born rule.

The effective Hamiltonian $\hat{H}$ has tiny quantum-gravitational
corrections, some of which render the Born rule unstable (Valentini 2021,
2023). The corrections are derived from (\ref{WD}) by a semiclassical
expansion of $\Psi$ (Kiefer and Singh 1991; Brizuela, Kiefer, and Kr\"{a}mer
2016). There are both Hermitian and non-Hermitian terms.\footnote{There is no
consensus as to whether the non-Hermitian terms are an artifact or a real
physical effect. Some authors advocate eliminating them by appropriate
redefinitions of the wave function (Kiefer and Wichmann 2018).} Applying the
same expansion to the de Broglie guidance equation (\ref{deB_QG}) for $\phi$,
the field velocity takes the same form as before but with $S$ equal to the
phase of $\psi$. Writing $\hat{H}=\hat{H}_{1}+i\hat{H}_{2}$ (with $\hat{H}%
_{1}$, $\hat{H}_{2}$ Hermitian), we then have the usual continuity equation
(\ref{Cont_rho_gen}) for $\rho$ (cf. (\ref{cont_QFICS})) while the continuity
equation (\ref{Cont_psi2_gen2}) for $|\psi|^{2}$ now has a source term
$s=2\operatorname{Re}\left(  \psi^{\ast}\hat{H}_{2}\psi\right)  $ on the
right-hand side. The continuity equations for $\rho$ and $\left\vert
\psi\right\vert ^{2}$ no longer match. Initial equilibrium $\rho=\left\vert
\psi\right\vert ^{2}$ can therefore evolve into final nonequilibrium $\rho
\neq\left\vert \psi\right\vert ^{2}$ on an estimated timescale%
\begin{equation}
\tau_{\mathrm{noneq}}\approx\frac{1}{2\left\vert \left\langle \hat{H}%
_{2}\right\rangle \right\vert }\ .
\end{equation}

For a scalar field in the vicinity of an evaporating Schwarzchild black hole
of mass $M(t)$, the Hamiltonian $\hat{H}_{\mathbf{k}}$ of a field mode has a
non-Hermitian correction $i\hat{H}_{2}$ with (Kiefer, M\"{u}ller, and Singh
1994)%
\begin{equation}
\hat{H}_{2}\simeq-\frac{1}{12}\kappa\left(  \frac{m_{\mathrm{P}}}{M}\right)
^{4}\hat{H}_{\mathbf{k}}\ ,
\end{equation}
where $\kappa$ is a numerical factor and here $m_{\mathrm{P}}=\sqrt{\hbar
c/G}\simeq10^{-5}\ \mathrm{g}$ is the standard Planck mass. We then find
(Valentini 2021, 2023)%
\begin{equation}
\tau_{\mathrm{noneq}}\sim\frac{48\pi}{\kappa}t_{\mathrm{P}}\left(  \frac
{M}{m_{\mathrm{P}}}\right)  ^{5}%
\end{equation}
(where $t_{\mathrm{P}}$ is the Planck time). The effect is significant when
$M$ approaches $m_{\mathrm{P}}$. The final burst of Hawking radiation could
violate the Born rule.\footnote{A breakdown of the Born rule in Hawking
radiation has also been suggested as a possible resolution of the
information-loss puzzle (Valentini 2004, 2007; Kandhadai and Valentini 2020).}
In principle such violations might be observed in radiation from exploding
primordial black holes.

\section{Conclusion}

The de Broglie-Bohm pilot-wave formulation of quantum mechanics avoids the
notorious quantum measurement problem while highlighting the nonlocality of
quantum physics. Quantum `measurements' are understood to be physical
processes like any other, obeying the same underlying physical laws. There is
no need for a fundamental division between `system' and `apparatus', and
physical processes may be described in objective terms even in the absence of
`observers'. Pilot-wave theory may be applied across the full range of known
physics, including high-energy physics, quantum field theory, gravitation, and
cosmology. It has proved to be especially useful in cosmology, an area where
textbook quantum mechanics is difficult to apply since there is no outside observer.

The internal logic of pilot-wave theory changes our perspective on the Born
rule, which can no longer be regarded as a law of physics. Instead it
corresponds to a state of statistical equilibrium, broadly analogous to
thermal equilibrium, which may be understood as having arisen by dynamical
relaxation. Quantum physics is seen to be merely a special case of a much
wider physics that includes nonlocal signalling, violations of the uncertainty
principle, and the breaking of other standard quantum constraints. Thus
pilot-wave theory provides a natural extension of quantum physics to a new
nonequilibrium regime.

Evidence for quantum nonequilibrium physics could be found in the early
universe, though this is still an active area of research with as yet no clear
conclusions. Relic nonequilibrium systems from early times might still exist
today. Exotic gravitational effects could potentially create nonequilibrium
where there was none before. Should nonequilibrium systems be discovered and
put to practical use, there would be radical technological implications,
including for cryptography and computing.

Pilot-wave theory is a distinctive approach to the theory of motion and to
physics generally. It provides a novel perspective on the structure of
spacetime, as well as on the foundations of quantum gravity and the nature of
black holes. While much remains to be understood, this formulation of quantum
mechanics continues to provide insights into the nature of quantum physics and
to suggest new lines of research that would not be available to other formulations.

\bigskip

\begin{center}
{\LARGE Bibliography}
\end{center}

\bigskip

Abraham, E., Colin, S., and Valentini, A. (2014). Long-time relaxation in
pilot-wave theory. \textit{Journal of Physics A} 47, 395306.

Abrams, D. S. and Lloyd, S. (1998). Nonlinear quantum mechanics implies
polynomial-time solution for NP-complete and \#P problems. \textit{Physical
Review Letters} 81, 3992--3995.

Aghanim, N. \textit{et al}. (Planck Collaboration) (2016). \textit{Planck}
2015 results. XI. CMB power spectra, likelihoods, and robustness of
parameters. \textit{Astronomy and Astrophysics} 594, A11.

Bacciagaluppi, G. (2003). Derivation of the symmetry postulates for identical
particles from pilot-wave theories. arXiv:quant-ph/0302099.

Bacciagaluppi, G. and Valentini, A. (2009). \textit{Quantum theory at the
crossroads: Reconsidering the 1927 Solvay conference}. Cambridge: Cambridge
University Press.

Bell, J. S. (1966). On the problem of hidden variables in quantum mechanics.
\textit{Reviews of Modern Physics} 38, 447--452.

Bell, J. S. (1986). Quantum field theory without observers. \textit{Physics
Reports} 137, 49--54.

Bell, J. S. (1987). \textit{Speakable and unspeakable in quantum mechanics}.
Cambridge: Cambridge University Press.

Bennett, C. H. (1992). Quantum cryptography using any two nonorthogonal
states. \textit{Physical Review Letters} 68, 3121--3124.

Berndl, K., D\"{u}rr, D., Goldstein, S., and Zangh\`{\i}, N. (1996).
Nonlocality, Lorentz invariance, and Bohmian quantum theory. \textit{Physical
Review A} 53, 2062--2073.

Bohm, D. (1952a). A suggested interpretation of the quantum theory in terms of
`hidden' variables. I. \textit{Physical Review} 85, 166--179.

Bohm, D. (1952b). A suggested interpretation of the quantum theory in terms of
`hidden' variables. II. \textit{Physical Review} 85, 180--193.

Bohm, D., Hiley, B. J., and Kaloyerou, P. N. (1987). An ontological basis for
the quantum theory. \textit{Physics Reports} 144, 321--375.

Bohm, D. and Hiley, B. J. (1993). \textit{The undivided universe: an
ontological interpretation of quantum theory}. London: Routledge.

Brizuela, D., Kiefer, C., and Kr\"{a}mer, M. (2016). Quantum-gravitational
effects on gauge-invariant scalar and tensor perturbations during inflation:
the de Sitter case. \textit{Physical Review D} 93, 104035.

Brown, L. M. (1958). Two-component fermion theory. \textit{Physical Review}
111, 957--964.

Colin, S. (2003). A deterministic Bell model. \textit{Physics Letters A} 317, 349--358.

Colin, S. (2012). Relaxation to quantum equilibrium for Dirac fermions in the
de Broglie-Bohm pilot-wave theory. \textit{Proceedings of the Royal Society A}
468, 1116--1135.

Colin, S. and Struyve, W. (2007). A Dirac sea pilot-wave model for quantum
field theory. \textit{Journal of Physics A} 40, 7309--7341.

Colin, S. and Valentini, A. (2013). Mechanism for the suppression of quantum
noise at large scales on expanding space. \textit{Physical Review D} 88, 103515.

Colin, S. and Valentini, A. (2014). Instability of quantum equilibrium in
Bohm's dynamics. \textit{Proceedings of the Royal Society A} 470, 20140288.

Colin, S. and Valentini, A. (2015). Primordial quantum nonequilibrium and
large-scale cosmic anomalies. \textit{Physical Review D} 92, 043520.

Colin, S. and Valentini, A. (2016). Robust predictions for the large-scale
cosmological power deficit from primordial quantum nonequilibrium.
\textit{International Journal of Modern Physics D} 25, 1650068.

de Broglie, L. (1928). La nouvelle dynamique des quanta. In
\textit{\'{E}lectrons et photons: Rapports et discussions du cinqui\`{e}me
conseil de physique}, pp 105--132. Paris: Gauthier-Villars. [English
translation: Bacciagaluppi, G. and Valentini, A. (2009).]

Drezet, A. (2021). Justifying Born's rule $P_{\alpha}=\left\vert \Psi_{\alpha
}\right\vert ^{2}$ using deterministic chaos, decoherence, and the de
Broglie-Bohm quantum theory. \textit{Entropy} 23, 1371.

D\"{u}rr, D., Goldstein, S., and Zangh\`{\i}, N. (1992). Quantum equilibrium
and the origin of absolute uncertainty. \textit{Journal of Statistical
Physics} 67, 843--907.

D\"{u}rr, D., Goldstein, S., M\"{u}nch-Berndl, K., and Zanghi, N. (1999).
Hypersurface Bohm-Dirac models. \textit{Physical Review A} 60, 2729--2736.

D\"{u}rr, D., Goldstein, S., Tumulka, R., and Zangh\`{\i}, N. (2004). Bohmian
mechanics and quantum field theory. \textit{Physical Review Letters} 93, 090402.

D\"{u}rr, D., Goldstein, S., Tumulka, R., and Zangh\`{\i}, N. (2005).
Bell-type quantum field theories. \textit{Journal of Physics A} 38, R1--R43.

D\"{u}rr, D. and Teufel, S. (2009). \textit{Bohmian mechanics: the physics and
mathematics of quantum theory}. Heidelberg: Springer-Verlag.

D\"{u}rr, D., Goldstein, S., Norsen, T., Struyve, W., and Zangh\`{\i}, N.
(2014). Can Bohmian mechanics be made relativistic? \textit{Proceedings of the
Royal Society A} 470, 20130699.

Efthymiopoulos, C., Contopoulos, G., and Tzemos, A. C. (2017). Chaos in de
Broglie-Bohm quantum mechanics and the dynamics of quantum relaxation.
\textit{Annales de la Fondation Louis de Broglie} 42, 133--159.

Ekert, A. K. (1991). Quantum cryptography based on Bell's theorem.
\textit{Physical Review Letters} 67, 661--663.

Hardy, L. (1992). Quantum mechanics, local realistic theories, and
Lorentz-invariant realistic theories. \textit{Physical Review Letters} 68, 2981--2984.

Holland, P. R. (1993). \textit{The quantum theory of motion: an account of the
de Broglie-Bohm causal interpretation of quantum mechanics}. Cambridge:
Cambridge University Press.

Ho\v{r}ava, P. (2009). Quantum gravity at a Lifshitz point. \textit{Physical
Review D} 79, 084008.

Horiguchi, T. (1994). Quantum potential interpretation of the Wheeler--DeWitt
equation. \textit{Modern Physics Letters A} 9, 1429--1443.

Isham, C. J. (1993). Canonical quantum gravity and the problem of time. In
Ibort, L. A. \& Rodriguez, M. A. (eds.) \textit{Integrable systems, quantum
groups, and quantum field theories}, pp 157--288. London: Kluwer.

Kaloyerou, P. N. (1994). The causal interpretation of the electromagnetic
field. \textit{Physics Reports} 244, 287--358.

Kandhadai, A. and Valentini, A. (2019). Perturbations and quantum relaxation.
Foundations of Physics 49, 1--23.

Kandhadai, A. and Valentini, A. (2020). Mechanism for nonlocal information
flow from black holes. \textit{International Journal of Modern Physics A} 35, 2050031.

Kiefer, C. (2012). \textit{Quantum gravity}. Oxford: Oxford University Press.

Kiefer, C., M\"{u}ller, R., and Singh, T. P. (1994). Quantum gravity and
non-unitarity in black hole evaporation. \textit{Modern Physics Letters A} 9, 2661--2670.

Kiefer, C. and Singh, T. P. (1991). Quantum gravitational corrections to the
functional Schr\"{o}dinger equation. \textit{Physical Review D} 44, 1067--1076.

Kiefer, C. and Wichmann, D. (2018). Semiclassical approximation of the
Wheeler-DeWitt equation: arbitrary orders and the question of unitarity.
\textit{General Relativity and Gravitation} 50, 66.

Kim, S. K., Namgung, W., Soh, K. S., and Yee, J. H. (1990). Equivalence
between the Weyl, Coulomb, and unitary gauges in the functional
Schr\"{o}dinger picture. \textit{Physical Review D} 41, 3792--3795.

Kosteleck\'{y}, V. A. and Mewes, M. (2002). Signals for Lorentz violation in
electrodynamics. \textit{Physical Review D} 66, 056005.

Kucha\v{r}, K. V. (2011). Time and interpretations of quantum gravity.
\textit{International Journal of Modern Physics D} 20, 3--86.

Lazarovici, D. and Hubert, M. (2019). How quantum mechanics can consistently
describe the use of itself. \textit{Scientific Reports} 9, 470.

Leibbrandt, G. (1987). Introduction to noncovariant gauges. \textit{Reviews of
Modern Physics} 59, 1067--1119.

Liddle, A. R. and Lyth, D. H. (2000). \textit{Cosmological inflation and
large-scale structure}. Cambridge: Cambridge University Press.

Lustosa, F. B., Colin, S., and Perez Bergliaffa, S. E. (2021). Quantum
relaxation in a system of harmonic oscillators with time-dependent coupling.
\textit{Proceedings of the Royal Society A} 477, 20200606.

Lustosa, F. B., Pinto-Neto, N., and Valentini, A. (2023). Evolution of quantum
non-equilibrium for coupled harmonic oscillators. \textit{Proceedings of the
Royal Society A} 479, 20220411.

Nielsen, M. A. and Chuang, I. L. (2000). \textit{Quantum computation and
quantum information}. Cambridge: Cambridge University Press.

Pinto-Neto, N. (2005). The Bohm interpretation of quantum cosmology.
\textit{Foundations of Physics} 35, 577--603.

Pinto-Neto, N. (2021). The de Broglie-Bohm quantum theory and its application
to quantum cosmology. \textit{Universe} 7, 134.

Pinto-Neto, N. and Fabris, J. C. (2013). Quantum cosmology from the de
Broglie--Bohm perspective. \textit{Classical and Quantum Gravity} 30, 143001.

Pinto-Neto, N. and Sergio Santini, E. (2002). The consistency of causal
quantum geometrodynamics and quantum field theory. \textit{General Relativity
and Gravitation} 34, 505--532.

Roser, P. and Valentini, A. (2014). Classical and quantum cosmology with York
time. \textit{Classical and Quantum Gravity} 31, 245001.

Sakurai, J. J. (1967). \textit{Advanced quantum mechanics}. Reading,
Massachusetts: Addison-Wesley.

Sebens, C. T. (2016). Constructing and constraining wave functions for
identical quantum particles. \textit{Studies in History and Philosophy of
Modern Physics} 56, 48--59.

Shtanov, Yu. V. (1996). Pilot wave quantum cosmology. \textit{Physical Review
D} 54, 2564--2570.

Struyve, W. and Valentini, A. (2009). De Broglie-Bohm guidance equations for
arbitrary Hamiltonians. \textit{Journal of Physics A} 42, 035301.

Towler, M. D., Russell, N. J., and Valentini, A. (2012). Time scales for
dynamical relaxation to the Born rule. \textit{Proceedings of the Royal
Society A} 468, 990--1013.

Tumulka, R. (2007). The `unromantic pictures' of quantum theory.\textit{
Journal of Physics A} 40, 3245--3273.

Underwood, N. G. (2018). Extreme quantum nonequilibrium, nodes, vorticity,
drift and relaxation retarding states. \textit{Journal of Physics A} 51, 055301.

Underwood, N. G. and Valentini, A. (2015). Quantum field theory of relic
nonequilibrium systems. \textit{Physical Review D} 92, 063531.

Underwood, N. G. and Valentini, A. (2020). Anomalous spectral lines and relic
quantum nonequilibrium. \textit{Physical Review D} 101, 043004.

Valentini, A. (1991a). Signal-locality, uncertainty, and the subquantum
H-theorem. I. \textit{Physics Letters A} 156, 5--11.

Valentini, A. (1991b). Signal-locality, uncertainty, and the subquantum
H-theorem, II. \textit{Physics Letters A} 158, 1--8.

Valentini, A. (1992). On the pilot-wave theory of classical, quantum and
subquantum physics. PhD thesis, International School for Advanced Studies,
Trieste, Italy. [http://hdl.handle.net/20.500.11767/4334]

Valentini, A. (1996). Pilot-wave theory of fields, gravitation and cosmology.
In Cushing, J. T., Fine, A. \& Goldstein, S. (eds.) \textit{Bohmian mechanics
and quantum theory: an appraisal}, pp 45--66. Dordrecht: Kluwer.

Valentini, A. (1997). On Galilean and Lorentz invariance in pilot-wave
dynamics. \textit{Physics Letters A} 228, 215--222.

Valentini, A. (2002a). Signal-locality and subquantum information in
deterministic hidden-variables theories. In Placek, T. \& Butterfield, J.
(eds.) \textit{Non-locality and modality}, pp 81--103. Dordrecht: Kluwer.

Valentini, A. (2002b). Signal-locality in hidden-variables theories.
\textit{Physics Letters A} 297, 273--278.

Valentini, A. (2002c). Subquantum information and computation.
\textit{Pramana---Journal of Physics} 59, 269--277.

Valentini, A. (2004). Black holes, information loss, and hidden variables. arXiv:hep-th/0407032.

Valentini, A. (2007). Astrophysical and cosmological tests of quantum theory.
\textit{Journal of Physics A} 40, 3285--3303.

Valentini, A. (2008). Hidden variables and the large-scale structure of
space-time. In Craig, W. L. \& Smith, Q. (eds.) \textit{Einstein, relativity
and absolute simultaneity}, pp 125--155. London: Routledge.

Valentini, A. (2010). Inflationary cosmology as a probe of primordial quantum
mechanics. \textit{Physical Review D} 82, 063513.

Valentini, A. (2020). Foundations of statistical mechanics and the status of
the Born rule in de Broglie-Bohm pilot-wave theory. In Allori, V. (ed.)
\textit{Statistical mechanics and scientific explanation: determinism,
indeterminism and laws of nature}, pp 423--477. Singapore: World Scientific.

Valentini, A. (2021). Quantum gravity and quantum probability. arXiv:2104.07966.

Valentini, A. (2023). Beyond the Born rule in quantum gravity.
\textit{Foundations of Physics} 53, 6.

Valentini, A. (2025). \textit{Introduction to quantum foundations and
pilot-wave theory}. Oxford: Oxford University Press.

Valentini, A. and Westman, H. (2005). Dynamical origin of quantum
probabilities. \textit{Proceedings of the Royal Society A} 461, 253--272.

Vigier, J. P. (1985). Nonlocal quantum potential interpretation of
relativistic actions at a distance in many-body problems. In Tarozzi, G. \&
van der Merwe, A. (eds.) \textit{Open questions in quantum physics: invited
papers on the foundations of microphysics}, pp 297--322. Dordrecht: Reidel.

Vink, J. C. (1992). Quantum potential interpretation of the wave function of
the universe. \textit{Nuclear Physics B} 369, 707--728.

Vitenti, S., Peter, P. and Valentini, A. (2019). Modeling the large-scale
power deficit with smooth and discontinuous primordial spectra.
\textit{Physical Review D} 100, 043506.

\end{document}